\documentclass[twocolumn, aps, superscriptaddress, prb, showpacs,floatfix,longbibliography]{revtex4-2}
\usepackage{dcolumn}
\usepackage{graphicx}
\usepackage{amsmath,amssymb}
\usepackage{bm,xcolor,url}
\usepackage{pict2e,hyperref}
\usepackage{here}

\begin{document}
\title {Geometric contribution to adiabatic amplification in non-Hermitian systems}

\author {Tomoki Ozawa}
\affiliation{Advanced Institute for Materials Research (WPI-AIMR), Tohoku University, Sendai 980-8577, Japan}
\author {Henning Schomerus}
\affiliation{Advanced Institute for Materials Research (WPI-AIMR), Tohoku University, Sendai 980-8577, Japan}
\affiliation{Department of Physics, Lancaster University, Lancaster LA1 4YB, United Kingdom}

\date{\today}
\begin{abstract}
Concepts from non-Hermitian quantum mechanics have proven useful in understanding and manipulating a variety of classical systems, such as those encountered in optics, classical mechanics, and metamaterial design. Recently, the non-Hermitian analog of the Berry phase for adiabatic processes was experimentally measured. 
In non-Hermitian systems, the Berry phase can have an imaginary part, which contributes to the amplification or decay of the total wave intensity. 
When the imaginary part of the Berry curvature is zero, this geometric amplification factor is determined solely by the initial and final points of the adiabatic path in  parameter space, and it does not depend on how these points are connected by the path. We list classes of non-Hermitian Hamiltonians where this path independence is guaranteed by suitable symmetries, and we find that, for some of these classes, the amplification factor can be written only in terms of the Petermann factors of the initial and final points. Our result can, in turn, be used to experimentally obtain the Petermann factor by observing how the norm of the wave function changes under adiabatic processes. We validate our theory using a couple of concrete examples of physical relevance. 
\end{abstract}

\maketitle

\section{Introduction}

Non-Hermitian quantum mechanics is a theoretical framework describing systems whose state vector, i.e., wavefunction, obeys a Schr\"odinger equation with a non-Hermitian Hamiltonian~\cite{Bender2007,Ashida2020}. Originally proposed to describe open, decaying, and statistical quantum systems, such as nuclei interacting with the environment~\cite{Gamow1928,Feshbach1958} or an array of superconducting vortices~\cite{Hatano1996,Hatano1997}, non-Hermitian quantum mechanics recently found applications in various \textit{classical} systems, and it has been rapidly developing since the realization of a $PT$-symmetric Hamiltonian in optics~\cite{Bender1998,Makris2008,Guo2009,Ruter2010,El-Ganainy2018}, and the subsequent generalization into non-Hermitian topological photonics~\cite{Poli:2015,Midya2018, Shen:2018, Kawabata:2019, Ota2020, Bergholtz:2021, Ding:2022, Okuma:2023}.

One distinguishing feature of non-Hermitian quantum mechanics is that the norm of the state vector is not conserved. Although the overall normalization of the state vector is not a measurable property in quantum systems, it is a physically meaningful property in classical setups; for example in optics the squared norm of the state vector describes the total intensity of the light in the system.
One origin of the change of the norm of the state vector is the imaginary part of the energy. Because of the non-Hermiticity of the Hamiltonian, the eigenenergies can have a nonzero imaginary part, which describes the transient amplification or decay of the state vector. Another origin of the change of the norm of the state vector is the complex non-Hermitian Berry’s phase~\cite{Garrison1988, Dattoli1990, Mondragon1996, Bliokh1999, Keck2003, Nesterov2008, Liang2013,Singhal2023,OzawaHayata}, which appears when one slowly (adiabatically) changes some parameters in the Hamiltonian. This contribution is geometric; the change of the norm of the state is determined by the path followed in  parameter space and the geometric properties of the state vector along the path, and it occurs even when all the instantaneous eigenenergies of the system along the path are real-valued. A further distinguishing feature of the non-Hermitian Berry’s phase is that it is gauge invariant, i.e., independent of the normalization of the basis vectors, even when the path is open. In contrast, the ordinary Berry's phase in Hermitian systems is gauge invariant only when the path is closed. This adiabatic amplification or decay of the state vector due to the non-Hermitian Berry’s phase has recently been experimentally probed in feedback-coupled mechanical oscillators~\cite{Singhal2023}.

In this paper, we show that, under some circumstances, the geometric contribution to the adiabatic amplification depends only on the initial and final points in the parameter space, and does not depend on the details of the path connecting the two points. The necessary and sufficient condition for this path-independent adiabatic amplification is the vanishing of the imaginary part of the appropriately defined Berry curvature. We identify several key cases in which this path independence holds, and we find that in some cases, such as reciprocal Hamiltonians, the adiabatic amplification factor is determined solely by the ratio of the Petermann factors at the initial and final points in the parameter space.

This provides a conceptual link to a previously unrelated concept from non-Hermitian physics. Just as the eigenenergies, the Petermann factor relates to the quasistationary properties of a fixed Hamiltonian, for which it characterizes the non-orthogonality associated with a given eigenstate \cite{Chalker:1998}. Originally, this factor was uncovered as a source of excess quantum noise \cite{Petermann:1979,Siegman:1989,Patra:2000}. Subsequently, it was found that it characterizes the general sensitivity of the system to perturbations \cite{Heiss:2000}, including dynamical perturbations in the form of external noise \cite{Schomerus2020}. The Petermann factor also highlights the unconventional physics near non-Hermitian degeneracies (exceptional points) \cite{Heiss:2000,Berry:2004,Miri2019}, where it diverges in accordance with a qualitatively altered, super-Lorentzian, lineshape \cite{Yoo:2011,Takata:2021,Hashemi:2022}, and it leads to the design of sensors with nonlinear response functions \cite{Wiersig:2014,Chen:2017,Wiersig2020}. However, determining the Petermann factor directly in experiments has been a significant challenge \cite{vanEijkelenborg:1996,Cheng:1996}.

Our results suggest that, in order to amplify the signal to a desired value, one does not need to precisely control the speed of the parameter change or the path in the parameter space. One just needs to identify the initial and final points in the parameter space which provide the desired amplification, and one needs to change the parameters slowly enough via the path connecting the two points that is most easily accessible in  a given experiment. When the amplification factor is determined solely by the Petermann factor, our results in turn provide a method to experimentally measure the Petermann factor directly through the geometric amplification of the signal.

Here is the outline of our paper. In Sec.~\ref{sec:amp}, we lay out a general framework of adiabatic amplification in non-Hermitian setups, and establish that a vanishing imaginary part of the Berry curvature is the necessary and sufficient condition for the adiabatic amplification to only depend on the properties of the initial and final points in the parameter space. In Sec.~\ref{sec:main_general}, we identify several key cases where the imaginary part of the Berry curvature becomes zero and hence the adiabatic amplification factor is only end-point dependent. In Sec.~\ref{sec:sum}, we give a summary of the general discussion given in Sec.~\ref{sec:main_general}. In Sec.~\ref{sec:models} we analyze two concrete models to examine the validity of the adiabatic theorem our formalism relies on. In particular, we first examine a paradigmatic two-level Hamiltonian, and then study a non-Hermitian lattice system of experimental relevance. Our conclusions, given in Sec.~\ref{sec:conclusions}, include a summary of the findings and an outlook on possible applications and further work.

\section{Adiabatic amplification}
\label{sec:amp}
We first review the geometric contribution to the intensity amplification in non-Hermitian systems. Consider a non-Hermitian Hamiltonian $H(\boldsymbol\lambda)$ that depends on a set of real parameters $\boldsymbol\lambda = (\lambda_1, \lambda_2, \cdots)$. 
The right and left eigenstates of a non-Hermitian Hamiltonian are generally different from each other, but they share the same eigenvalues. We write the right and left eigenstates with an eigenvalue $E(\boldsymbol\lambda)$ as $H(\boldsymbol\lambda) |R(\boldsymbol\lambda)\rangle = E(\boldsymbol\lambda)|R(\boldsymbol\lambda)\rangle$ and $\langle L(\boldsymbol\lambda)| H(\boldsymbol\lambda) = E(\boldsymbol\lambda)\langle L(\boldsymbol\lambda)|$, respectively.
Throughout this paper, we do not assume any particular normalization of the right and left eigenstates; we allow $\langle R(\boldsymbol\lambda)|R(\boldsymbol\lambda)\rangle$, $\langle L(\boldsymbol\lambda)|R(\boldsymbol\lambda)\rangle$, and $\langle L(\boldsymbol\lambda)|L(\boldsymbol\lambda)\rangle$ to take arbitrary nonzero values.

We consider the dynamical evolution of the system where we start from a right eigenstate of the Hamiltonian at some point $\boldsymbol\lambda(0)$ in the parameter space,  and then adiabatically change the parameters as $\boldsymbol\lambda (t)$ until some time $t=T$. The evolution of the wavefunction $|\psi (t)\rangle$ follows the Schrödinger equation
\begin{align}
    i\frac{d}{dt}|\psi (t)\rangle = H(\boldsymbol\lambda (t))|\psi(t)\rangle. \label{eq:sch}
\end{align}
The initial condition is $|\psi (0)\rangle = c |R(\boldsymbol\lambda(0))\rangle$ with a constant $c$ determined by the initial norm of the wavefunction. We will examine the change of the intensity $I(t) \equiv \langle \psi (t)|\psi (t)\rangle$, which we will separate into a dynamical and a geometric contribution. 

The assumption of adiabaticity in Hermitian systems implies that the wavefunction remains proportional to the instantaneous eigenstate during the evolution, which is the consequence of the \textit{adiabatic theorem}. 
In non-Hermitian setups, for the adiabatic theorem to hold, it is required that not only the change of the parameters $\boldsymbol\lambda(t)$ is slow compared to the scale determined by the energy gap, but also that the eigenvalue corresponding to the right eigenstate $|R(\boldsymbol\lambda (t))\rangle$ has the largest imaginary part among all the eigenvalues of the Hamiltonian $H(\boldsymbol\lambda (t))$ at any time $t$~\cite{Nenciu:1992,Holler:2020,Graefe:2013}.
Under these conditions, in non-Hermitian systems, the wavefunction remains proportional to the instantaneous right eigenstate $|R(\boldsymbol\lambda (t))\rangle$ during the evolution.
An important case where this adiabaticity can hold is when all the eigenvalues are real, such as in the $PT$-symmetric regions in $PT$-symmetric Hamiltonians~\cite{Bender1998,Makris2008,Guo2009,Ruter2010,El-Ganainy2018}. 

Assuming that the adiabatic theorem holds, the Schr\"odinger equation (\ref{eq:sch}) can be integrated, as explained in Ref.~\cite{Singhal2023}, and the wavefunction at the final time $T$ can be written as
\begin{align}
    &|\psi (T)\rangle =
    \notag \\
    &c \cdot \exp
    \left[ - i \int_0^T E(\boldsymbol\lambda (t))dt + i\int_{\mathcal{C}} \boldsymbol{\mathcal{A}}^{LR} (\boldsymbol\lambda)\cdot d\boldsymbol\lambda \right]
    |R (\boldsymbol\lambda(T))\rangle, \label{eq:integrated}
\end{align}
where we have defined the left-right Berry connection by
\begin{align}
    \boldsymbol{\mathcal{A}}^{LR} (\boldsymbol\lambda)
    =
    i\frac{\langle L (\boldsymbol\lambda) | \nabla_{\boldsymbol\lambda} R(\boldsymbol\lambda)\rangle}{\langle L (\boldsymbol\lambda)|R(\boldsymbol\lambda)\rangle}. \label{eq:alr}
\end{align}
We will later also use another Berry connection, $\boldsymbol{\mathcal{A}}^{RR} (\boldsymbol\lambda)$, which we define by
\begin{align}
    \boldsymbol{\mathcal{A}}^{RR} (\boldsymbol\lambda)
    =
    i\frac{\langle R (\boldsymbol\lambda) | \nabla_{\boldsymbol\lambda} R(\boldsymbol\lambda)\rangle}{\langle R (\boldsymbol\lambda)|R(\boldsymbol\lambda)\rangle}. \label{eq:arr}
\end{align}
The integral of the left-right Berry connection in Eq.~(\ref{eq:integrated}) is the line integral along the path $\mathcal{C}$ in the parameter space $\boldsymbol\lambda$ drawn by $\boldsymbol\lambda(t)$.

The ratio of the squared norm (total intensity) of the wavefunction at the final time to the initial time, which we call the adiabatic amplification factor, is
\begin{align}
    &\frac{I(T)}{I(0)}
    =
    \frac{\langle \psi (T)|\psi(T)\rangle}{\langle \psi (0)|\psi(0)\rangle}
    =
    \frac{\langle R (\boldsymbol\lambda (T))|R(\boldsymbol\lambda (T))\rangle}{\langle R (\boldsymbol\lambda (0))|R(\boldsymbol\lambda (0))\rangle}
    \times
    \notag \\
    &\exp
    \left[ 2 \int_0^T \mathrm{Im}\left[E(\boldsymbol\lambda (t))\right]dt - 2\int_{\mathcal{C}} \mathrm{Im}\left[\boldsymbol{\mathcal{A}}^{LR} (\boldsymbol\lambda)\right]\cdot d\boldsymbol\lambda \right].
\end{align}
We can further show that
\begin{align}
    \frac{\langle R (\boldsymbol\lambda (T))|R(\boldsymbol\lambda (T))\rangle}{\langle R (\boldsymbol\lambda (0))|R(\boldsymbol\lambda (0))\rangle}
    =
    \exp \left[2\int_\mathcal{C} \mathrm{Im}\left[ \boldsymbol{\mathcal{A}}^{RR} (\boldsymbol\lambda)\right]\cdot d\boldsymbol\lambda \right],
\end{align}
which follows from integrating
\begin{align}
    \nabla_{\boldsymbol\lambda} \log \left( \langle R (\boldsymbol\lambda)|R(\boldsymbol\lambda)\rangle \right)
    =
    2\,
    \mathrm{Im}\left[ \boldsymbol{\mathcal{A}}^{RR} (\boldsymbol\lambda)\right].
\end{align}
Then, the adiabatic amplification factor is
\begin{align}
    \frac{I(T)}{I(0)}
    =&
    \exp
    \left[ 2 \int_0^T \mathrm{Im}\left[E(\boldsymbol\lambda (t))\right]dt \right]\times
    \notag \\
    &\exp
    \left[ - 2\int_{\mathcal{C}} \mathrm{Im}\left[ \boldsymbol{\mathcal{A}}^{LR} (\boldsymbol\lambda) - \boldsymbol{\mathcal{A}}^{RR} (\boldsymbol\lambda)\right]\cdot d\boldsymbol\lambda \right].
\end{align}
The first factor, which involves the imaginary part of the eigenvalue, is a dynamical factor that depends on the speed with which one is changing the parameters. The second factor, which only depends on the path $\mathcal{C}$ in the parameter space and does not depend on the speed of the parameter change, is a geometric contribution to the adiabatic amplification factor.
We note that, as discussed in earlier works~\cite{Silberstein2020,Singhal2023}, the combination $\boldsymbol{\mathcal{A}}^{LR} (\boldsymbol\lambda) - \boldsymbol{\mathcal{A}}^{RR}(\boldsymbol\lambda)$, which appears in the geometric contribution, is invariant under the (generalized) gauge transformation to change $|R(\boldsymbol\lambda)\rangle$ and $|L(\boldsymbol\lambda)\rangle$ by a nonzero $\boldsymbol\lambda$-dependent multiplicative factor. In other words, $\boldsymbol{\mathcal{A}}^{LR} (\boldsymbol\lambda) - \boldsymbol{\mathcal{A}}^{RR}(\boldsymbol\lambda)$ is independent of the normalization of the right and left eigenstates.

When the eigenvalues of the Hamiltonian are all real, the adiabatic amplification factor is determined solely by the geometric term.
When the eigenvalue $E(\boldsymbol\lambda(t))$ is a complex number with nonzero imaginary part, the adiabatic amplification factor generally contains the dynamical factor. However, we note that we can cancel the dynamical factor and single out the geometric factor if we also consider the adiabatic amplification factor of the inverse process, where the parameters change adiabatically as $\boldsymbol\lambda (T - t)$, and take the ratio of the adiabatic amplification factors for these two processes.

The geometric contribution to the adiabatic amplification factor, which we denote by
\begin{align}
    A_g (\mathcal{C}) 
    &= 
    \left.\frac{I(T)}{I(0)}\right|_{\mathrm{geometric}}
    \notag \\
    &\equiv
    \exp
    \left[ - 2\int_{\mathcal{C}} \mathrm{Im}\left[ \boldsymbol{\mathcal{A}}^{LR} (\boldsymbol\lambda) - \boldsymbol{\mathcal{A}}^{RR} (\boldsymbol\lambda)\right]\cdot d\boldsymbol\lambda \right], \label{eq:defgaf}
\end{align}
only depends on the path $\mathcal{C}$ in the parameter space.

Our primary interest is to understand under which circumstances this geometric amplification factor only depends on the choice of initial and final points of the adiabatic process in parameter space, and not on the specific path that connects the chosen two points. 

One obvious case is when the parameter space is one dimensional. The adiabatic amplification factor is then determined by the values of the primitive function of the integrand at the initial and final points of the path. 

We therefore consider the non-trivial case of the parameter space being two dimensional or higher, and we study the conditions under which the geometric contribution to the adiabatic amplification factor is only determined by the initial and final points of the path. 

Let us consider two different paths, $\mathcal{C}$ and $\mathcal{C}^\prime$, which share the same initial and final points in the parameter space. The condition that the two paths should give the same geometric amplification factor is equivalent to
\begin{align}
    \exp
    \left[ - 2\int_{\mathcal{C} - \mathcal{C}^\prime} \mathrm{Im}\left[ \boldsymbol{\mathcal{A}}^{LR} (\boldsymbol\lambda) - \boldsymbol{\mathcal{A}}^{RR} (\boldsymbol\lambda)\right]\cdot d\boldsymbol\lambda \right]
    =
    1, \label{eq:cond}
\end{align}
where the path $\mathcal{C} - \mathcal{C}^\prime$ is the closed path obtained by joining the two paths so that path $\mathcal{C}$ is followed by path $\mathcal{C}^\prime$ in the backward direction. Let us introduce the Berry curvature to rewrite Eq.~(\ref{eq:cond}) in terms of a surface integral when the path $\mathcal{C} - \mathcal{C}^\prime$ is contractible to a point. If $\boldsymbol\lambda$ is three dimensional~
\footnote{If $\boldsymbol\lambda$ is not three-dimensional, we need to define the Berry curvature as a two-form via the Berry connection one-form by
\begin{align}
    \sum_{i<j}\Omega^{LR}_{ij} (\boldsymbol\lambda) d\lambda_i \wedge d\lambda_j \equiv \sum_{i,j} \partial_{\lambda_i}\mathcal{A}^{LR}_j (\boldsymbol\lambda) d\lambda_i \wedge d\lambda_j.
\end{align}},
we define the left-right and right-right Berry curvatures by
\begin{align}
    \boldsymbol\Omega^{LR} (\boldsymbol\lambda) \equiv \nabla_{\boldsymbol\lambda} \times \boldsymbol{\mathcal{A}}^{LR} (\boldsymbol\lambda), \notag \\
    \boldsymbol\Omega^{RR} (\boldsymbol\lambda) \equiv \nabla_{\boldsymbol\lambda} \times \boldsymbol{\mathcal{A}}^{RR} (\boldsymbol\lambda).
\end{align}
We note that $\mathrm{Im}\left[\boldsymbol{\Omega}^{RR} (\boldsymbol\lambda)\right] = 0$ always holds, but $\mathrm{Im}\left[\boldsymbol{\Omega}^{LR} (\boldsymbol\lambda)\right]$ is generally nonzero.
Then, Eq.~(\ref{eq:cond}) can be written as
\begin{align}
    \exp
    \left[ - 2\int_{\mathcal{S}} \mathrm{Im}\left[ \boldsymbol{\Omega}^{LR} (\boldsymbol\lambda)\right]\cdot d\hat{n} \right]
    =
    1, \label{eq:cond2}
\end{align}
for a contractible path $\mathcal{C} - \mathcal{C}^\prime$, where the integral is a surface integral over any surface $\mathcal{S}$ enclosed by the path $\mathcal{C} - \mathcal{C}^\prime$, and  $d\hat{n}$ is the normal vector perpendicular to $\mathcal{S}$.
Imposing the condition Eq.~(\ref{eq:cond2}) for any paths $\mathcal{C}$ and $\mathcal{C}^\prime$, where $\mathcal{C} - \mathcal{C}^\prime$ is contractible, and any surface $\mathcal{S}$ enclosed by the path $\mathcal{C} - \mathcal{C}^\prime$, we arrive at the condition
\begin{align}
    \mathrm{Im}\left[ \boldsymbol{\Omega}^{LR} (\boldsymbol\lambda)\right] =
    0, \label{eq:cond3}
\end{align}
under which the geometric amplification factor is invariant under the continuous change of the path. 

We make a remark here regarding the condition that $\mathcal{C} - \mathcal{C}^\prime$ be contractible. This condition implies that the geometric amplification factor can be different between $\mathcal{C}$ and $\mathcal{C}^\prime$ if the path $\mathcal{C} - \mathcal{C}^\prime$ forms a nontrivial loop in the parameter space and thus we cannot write the surface integral Eq.~(\ref{eq:cond2}). The contractibility condition is always fulfilled when the parameter space is simply connected. For simply connected parameter spaces, the condition Eq.~(\ref{eq:cond3}) is the necessary and sufficient condition for the geometric amplification factor to depend only on the choice of the initial and final points in the parameter space. In the rest of the paper, we assume that the parameter space is simply connected.

In the next section, we identify key cases where the condition Eq.~(\ref{eq:cond3}) holds, and derive expressions of the geometric amplification factor in terms of local properties at the initial and final points in  parameter space.

\section{Path-independent formulation of the geometric amplification factor}
\label{sec:main_general}

We now analyze the geometric amplification factor, Eq.~(\ref{eq:defgaf}), under the condition Eq.~(\ref{eq:cond3}). 
We find that this condition
is fulfilled in the following two classes of systems:
\begin{enumerate}
    \item[A.] When there exists a $\boldsymbol\lambda$-independent Hermitian matrix $M$ such that $|L(\boldsymbol\lambda)\rangle = M|R(\boldsymbol\lambda)\rangle$
    \item[B.] When there exists a $\boldsymbol\lambda$-independent symmetric matrix $M$ such that $|L(\boldsymbol\lambda)\rangle = M|R(\boldsymbol\lambda)^*\rangle$
\end{enumerate}
and also in two similar classes:
\begin{enumerate}
    \item[A$^\prime$.] When there exists a $\boldsymbol\lambda$-independent Hermitian matrix $M$ such that $|R(\boldsymbol\lambda)\rangle = M|L(\boldsymbol\lambda)\rangle$
    \item[B$^\prime$.] When there exists a $\boldsymbol\lambda$-independent symmetric matrix $M$ such that $|R(\boldsymbol\lambda)\rangle = M|L(\boldsymbol\lambda)^*\rangle$
\end{enumerate}
Here, $|R(\boldsymbol\lambda)^*\rangle$ and $|L(\boldsymbol\lambda)^*\rangle$ are the complex conjugates of $|R(\boldsymbol\lambda)\rangle$ and $|L(\boldsymbol\lambda)\rangle$, respectively.
One can check that, for the cases B and B$^\prime$, not only the imaginary part, but also the real part of the Berry curvature vanishes: $ \boldsymbol{\Omega}^{LR} (\boldsymbol\lambda) = 0$.

The cases A. and A$^\prime$. are equivalent when $M$ is invertible, and so are the cases B. and B$^\prime$. However we do not need to impose the invertibility of the matrix $M$; this is the reason we need to consider these cases separately.
An important example of a non-invertible $M$ is when $M$ is a rank-1 projector, in which case $|L(\boldsymbol\lambda)\rangle = M|R(\boldsymbol\lambda)\rangle$ implies that $|L(\boldsymbol\lambda)\rangle$ can be chosen to be a $\boldsymbol\lambda$-independent vector; we will come back to this example later.

Before examining each case to find the path-independent formula for the geometric amplification factor, we derive some useful relations about the integrand of the geometric amplification factor.
First, let us note that the geometric amplification factor can be written also in terms of the projectors
\begin{align}
    P_L (\boldsymbol\lambda) &\equiv \frac{|L(\boldsymbol\lambda)\rangle\langle L(\boldsymbol\lambda)|}{\langle L (\boldsymbol\lambda)|L(\boldsymbol\lambda)\rangle},
    &
    P_R (\boldsymbol\lambda) &\equiv \frac{|R(\boldsymbol\lambda)\rangle\langle R(\boldsymbol\lambda)|}{\langle R (\boldsymbol\lambda)|R(\boldsymbol\lambda)\rangle},
\end{align}
as
\begin{align}
    A_g (\mathcal{C})
    =
    \exp
    \left[- \int_{\mathcal{C}} \frac{\mathrm{Tr}\left( P_L \nabla_{\boldsymbol\lambda} P_R\right)}{\mathrm{Tr}\left( P_L P_R\right)}\cdot d\boldsymbol\lambda \right]. \label{eq:geoinp}
\end{align}
We also introduce here the Petermann factor, which is a local gauge-invariant quantity characterizing the different orientation of the right and left eigenstates, and it is defined by
\begin{align}
    K(\boldsymbol\lambda) \equiv \frac{\langle L(\boldsymbol\lambda)|L(\boldsymbol\lambda)\rangle \langle R(\boldsymbol\lambda)|R(\boldsymbol\lambda)\rangle}{|\langle L(\boldsymbol\lambda) |R(\boldsymbol\lambda)\rangle|^2}
    =
    \frac{1}{\mathrm{Tr}(P_L P_R)}. \label{eq:petermann}
\end{align}
We will see shortly that, in many cases, the geometric amplification factor can be written in terms of the Petermann factors at the initial and the final points in the parameter space. 

The integrand of Eq.~(\ref{eq:geoinp}) can be written as
\begin{align}
    &\frac{\mathrm{Tr}\left( P_L \nabla_{\boldsymbol\lambda} P_R\right)}{\mathrm{Tr}\left( P_L P_R\right)}
    =
    \nabla_{\boldsymbol\lambda}\left( \ln \frac{1}{\langle R|R\rangle}\right) + \left( \frac{\langle L | \nabla_{\boldsymbol\lambda}R\rangle}{\langle L | R\rangle} + c.c.\right), \label{eq:useful}
\end{align}
where $c.c.$ stands for the complex conjugate of the term within the same parenthesis,
and similarly
\begin{align}
    \frac{\mathrm{Tr}\left( (\nabla_{\boldsymbol\lambda} P_L)P_R\right)}{\mathrm{Tr}\left( P_L P_R\right)}
    =
    \nabla_{\boldsymbol\lambda}\left( \ln \frac{1}{\langle L|L\rangle}\right) + \left( \frac{\langle R | \nabla_{\boldsymbol\lambda}L\rangle}{\langle R | L\rangle} + c.c.\right). \label{eq:useful2}
\end{align}
Since the integrand of Eq.~(\ref{eq:geoinp}) can also be written as
\begin{align}
    \frac{\mathrm{Tr}\left( P_L \nabla_{\boldsymbol\lambda} P_R\right)}{\mathrm{Tr}\left( P_L P_R\right)}
    =
    \frac{\nabla_{\boldsymbol\lambda}\mathrm{Tr}\left( P_L P_R\right)}{\mathrm{Tr}\left( P_L P_R\right)} - \frac{\mathrm{Tr}\left( (\nabla_{\boldsymbol\lambda} P_L)P_R\right)}{\mathrm{Tr}\left( P_L P_R\right)}, \label{eq:useful3}
\end{align}
adding Eqs.~(\ref{eq:useful3}) and~(\ref{eq:useful}), using Eq.~~(\ref{eq:useful2}), and dividing by 2, we obtain that the integrand is
\begin{align}
    \frac{\mathrm{Tr}\left( P_L \nabla_{\boldsymbol\lambda} P_R\right)}{\mathrm{Tr}\left( P_L P_R\right)}
    &=
    \frac{1}{2}\frac{\nabla_{\boldsymbol\lambda}\mathrm{Tr}\left( P_L P_R\right)}{\mathrm{Tr}\left( P_L P_R\right)} 
    +
    \frac{1}{2}\nabla_{\boldsymbol\lambda}\left( \ln \frac{\langle L|L\rangle}{\langle R|R\rangle}\right)
    \notag \\
    +
    \frac{1}{2}&\left( \frac{\langle L | \nabla_{\boldsymbol\lambda}R\rangle}{\langle L | R\rangle} - \frac{\langle R | \nabla_{\boldsymbol\lambda}L\rangle}{\langle R | L\rangle} + c.c. \right)
    \notag \\
    &= \nabla_{\boldsymbol\lambda} \ln \sqrt{\frac{1}{K}\frac{\langle L|L\rangle}{\langle R|R\rangle}} + \Xi (\boldsymbol\lambda),
\end{align}
where we have grouped some terms as
\begin{align}
    \boldsymbol\Xi (\boldsymbol\lambda) &\equiv \frac{1}{2}\left( \frac{\langle L | \nabla_{\boldsymbol\lambda}R\rangle}{\langle L | R\rangle} - \frac{\langle R | \nabla_{\boldsymbol\lambda}L\rangle}{\langle R | L\rangle} + c.c. \right)
    \notag \\
    &= \mathrm{Im}\left[ \boldsymbol{\mathcal{A}}^{LR}(\boldsymbol\lambda) - \boldsymbol{\mathcal{A}}^{RL}(\boldsymbol\lambda) \right]
\end{align}
The geometric amplification factor can then be written as
\begin{align}
    A_g (\mathcal{C})
    =&
    \sqrt{K_T \frac{\langle R_T|R_T\rangle}{\langle L_T | L_T\rangle} \frac{1}{K_0}\frac{\langle L_0|L_0\rangle}{\langle R_0|R_0\rangle}}
    \times\notag \\
    &\exp \left[- \int_{\mathcal{C}} \boldsymbol\Xi (\boldsymbol\lambda) \cdot d\boldsymbol\lambda \right].
\end{align}
Here, $|R_0\rangle$ and $|L_0\rangle$ are the right and left eigenstates of the Hamiltonian at $t = 0$, and $|R_T\rangle$ and $|L_T\rangle$ are those at $t = T$.
As one can see, the square-root factor is solely determined by the properties at the initial and final points of the path $\mathcal{C}$ in parameter space, but the integral of $\boldsymbol\Xi (\boldsymbol\lambda)$ still depends on the path $\mathcal{C}$ itself. 

When $\boldsymbol\Xi (\boldsymbol\lambda) = 0$, it follows that $\nabla_{\boldsymbol\lambda} \times \boldsymbol\Xi (\boldsymbol\lambda) = \mathrm{Im}\left[ \boldsymbol{\Omega}^{LR} (\boldsymbol\lambda) - \boldsymbol{\Omega}^{RL} (\boldsymbol\lambda)\right] = 2\, \mathrm{Im}\left[ \boldsymbol{\Omega}^{LR} (\boldsymbol\lambda)\right] = 0$. (We have used $\mathrm{Im}\left[ \boldsymbol{\Omega}^{LR} (\boldsymbol\lambda)\right] = - \mathrm{Im}\left[ \boldsymbol{\Omega}^{RL} (\boldsymbol\lambda)\right]$). Since $\mathrm{Im}\left[ \boldsymbol{\Omega}^{LR} (\boldsymbol\lambda)\right] = 0$, as discussed at Eq.~(\ref{eq:cond3}), is the necessary and sufficient condition to have the geometric amplification factor $A_g (\mathcal{C})$ to be path independent, the condition $\boldsymbol\Xi (\boldsymbol\lambda) = 0$ provides a sufficient condition for the path independence. We note that the converse is not always true; $\mathrm{Im}\left[ \boldsymbol{\Omega}^{LR} (\boldsymbol\lambda)\right] = 0$ does not imply $\boldsymbol\Xi (\boldsymbol\lambda) = 0$.

As we will see, when one of the conditions we listed in the beginning of this section is met, $\boldsymbol\Xi (\boldsymbol\lambda) = 0$ holds, and thus $\mathrm{Im}\left[ \boldsymbol{\Omega}^{LR} (\boldsymbol\lambda)\right] = 0$ also holds. We now consider each of these cases separately.

\subsection{$|L\rangle = M|R\rangle$ with a Hermitian $M$}

When the right and left eigenstates are linearly related by a Hermitian matrix $M$ as $|L\rangle = M|R\rangle$, we see that
\begin{align}
    \boldsymbol\Xi (\boldsymbol\lambda) &= \frac{1}{2}\left( \frac{\langle R | M^\dagger \nabla_{\boldsymbol\lambda}R\rangle}{\langle R | M^\dagger|R\rangle} - \frac{\langle R | \nabla_{\boldsymbol\lambda} MR\rangle}{\langle R |M| R\rangle} + c.c. \right) \notag \\
    &= 0.
\end{align}
Thus, the geometric amplification factor is
\begin{align}
    A_g (\mathcal{C})
    &=
    \sqrt{K_T \frac{\langle R_T|R_T\rangle}{\langle L_T | L_T\rangle}\cdot \frac{1}{K_0}\frac{\langle L_0|L_0\rangle}{\langle R_0|R_0\rangle}}
    \notag \\
    &=
    \sqrt{K_T \frac{\langle R_T|R_T\rangle}{\langle R_T |M^2| R_T\rangle}\cdot \frac{1}{K_0}\frac{\langle R_0|M^2|R_0\rangle}{\langle R_0|R_0\rangle}}. \label{eq:gafa}
\end{align}
We note that the final expression is manifestly invariant under the change of the norm and/or phase of the right eigenstates at the initial and the final positions.

One noticeable example in which $|L\rangle = M|R\rangle$ with a Hermitian $M$ holds is when the non-Hermitian Hamiltonian $H(\boldsymbol\lambda)$ can be transformed into a Hermitian matrix under a similarity transformation---namely, if there exists an invertible $\boldsymbol\lambda$-independent matrix $P$ with 
\begin{align}
    \left( P^{-1}H(\boldsymbol\lambda) P \right)^\dagger = P^{-1}H(\boldsymbol\lambda) P.
\end{align}
One can then show that the eigenvalues of $H(\boldsymbol\lambda)$ are all real, and $|L\rangle = M|R\rangle$ holds with $M = (P P^{\dagger})^{-1}$, which is a Hermitian matrix.

We now discuss two special cases, where $M$ is also a unitary matrix, and where $M$ is a projector. As we see, for these two cases, the geometric amplification factor is written solely in terms of the Petermann factors at the initial and final points in the parameter space. 

\subsubsection{When $M$ is unitary}
When $M$ is unitary as well as Hermitian, $M^2 = I$ holds, where $I$ is the identity matrix. Important nontrivial examples where $M$ is both Hermitian and unitary are the Pauli matrices and any tensor products of them. The geometric amplification factor then reduces to 
\begin{align}
    A_g (\mathcal{C})
    &=
    \sqrt{\frac{K_T}{K_0}}, \label{eq:mhermunit}
\end{align}
which is a particularly simple form written solely in terms of the Petermann factors at the initial and final points in the parameter space.

\subsubsection{When $M$ is a projector}
When $M$ is a projector, $M^2 = M$ holds, and we see the following relations:
\begin{align}
    \langle L|L\rangle = \langle R|M^2|R\rangle = \langle R|M|R\rangle = \langle L|R\rangle = \langle R|L\rangle.
\end{align}
Then,
\begin{align}
    \frac{\langle R|R\rangle}{\langle L|L\rangle} = \frac{\langle R|R\rangle \langle L|L\rangle}{|\langle L|R\rangle|^2} = K.
    \label{eq:rrllk}
\end{align}
Using Eq.~(\ref{eq:rrllk}) in the first line of Eq.~(\ref{eq:gafa}), we obtain
\begin{align}
    A_g (\mathcal{C})
    &=
    \frac{K_T}{K_0},
\end{align}
and thus the geometric amplification factor is again written solely in terms of the Petermann factors,but now we have a linear dependence, instead of the square-root dependence we saw in Eq.~(\ref{eq:mhermunit}).

An important example where $M$ is both Hermitian and a projector is when $|L\rangle$ can be taken independent of $\boldsymbol\lambda$. We can then choose $M = |L\rangle \langle L|$ with a normalized $|L\rangle$, and we see that $|L\rangle \propto M|R\rangle$~\footnote{We note that, since $\langle L | R\rangle \neq 0$ for left and right eigenstates of the same nondegenerate eigenvalue, $M|R\rangle$ never vanishes.}. In this case, $M$ is a rank-1 projector. We will later consider a model where $|L\rangle$ is $\boldsymbol\lambda$-independent, and verify by numerical simulations that the geometric amplification factor is indeed given by $K_T/K_0$.

\subsection{$|L\rangle = M|R^*\rangle$ with a symmetric $M$}

When the right and left eigenstates are anti-linearly related by a symmetric matrix $M$ as $|L\rangle = M|R^*\rangle$, we see that
\begin{align}
    \boldsymbol\Xi (\boldsymbol\lambda) &= \frac{1}{2}\left( \frac{\langle R^* | M^* \nabla_{\boldsymbol\lambda}R\rangle}{\langle R^* | M^*|R\rangle} - \frac{\langle R | \nabla_{\boldsymbol\lambda} MR^*\rangle}{\langle R |M| R^*\rangle} + c.c. \right).
\end{align}
Since the first term in the parenthesis is the complex conjugate of the second term, the first and second terms give a purely imaginary value. Therefore, adding its complex conjugate, we see that $\boldsymbol\Xi (\boldsymbol\lambda) = 0$, and the geometric amplification factor is
\begin{align}
    &A_g (\mathcal{C})
    =
    \sqrt{K_T \frac{\langle R_T|R_T\rangle}{\langle L_T | L_T\rangle}\cdot \frac{1}{K_0}\frac{\langle L_0|L_0\rangle}{\langle R_0|R_0\rangle}}
    \notag \\
    &=
    \sqrt{K_T \frac{\langle R_T|R_T\rangle}{\langle R_T^* |M^* M| R_T^*\rangle}\cdot \frac{1}{K_0}\frac{\langle R_0^*|M^* M|R_0^*\rangle}{\langle R_0|R_0\rangle}}. \label{eq:gafa2}
\end{align}
This expression is, again, manifestly invariant under the change of the norm and/or phase of the right eigenstates at the initial and final points in the parameter space.

One noticeable example in which $|L\rangle = M|R^*\rangle$ with a symmetric $M$ holds is when the non-Hermitian Hamiltonian $H(\boldsymbol\lambda)$ can be transformed into a symmetric matrix under a similarity transformation---namely, if there exists an invertible $\boldsymbol\lambda$-independent matrix $P$ with 
\begin{align}
    \left( P^{-1}H(\boldsymbol\lambda) P \right)^T = P^{-1}H(\boldsymbol\lambda) P. \label{eq:php_sym}
\end{align}
One can then show that $|L\rangle = M|R^*\rangle$ holds with $M = (P^* P^{\dagger})^{-1}$, which is a symmetric matrix.
In particular, when the Hamiltonian $H(\boldsymbol\lambda)$ is symmetric, which constitutes an important class of non-Hermitian Hamiltonians representing reciprocal systems, the condition $|L\rangle = |R^*\rangle$ holds with $M = I$.

We again consider two special cases, where $M$ is also a unitary matrix, and where $M$ is a rank-1 projector.

\subsubsection{When $M$ is unitary}
\label{sec:reciprocal}
When $M$ is unitary as well as symmetric, $M M^* = I$ holds.
An important example of $M$ being unitary is when the Hamiltonian can be 
transformed into a reciprocal (symmetric) form with a fixed unitary matrix---namely, when $P$ in Eq.~(\ref{eq:php_sym}) is a unitary matrix, and thus $M = (P^* P^{\dagger})^{-1}$ is unitary as well.

Noting that $\langle R^*|R^*\rangle = \langle R|R\rangle$ generally holds, the geometric amplification factor when $M$ is unitary is
\begin{align}
    A_g (\mathcal{C})
    &=
    \sqrt{\frac{K_T}{K_0}},
\end{align}
which is again written solely in terms of the Petermann factors at the initial and final points in the parameter space. We note that this result is the same as Eq.~(\ref{eq:mhermunit}), namely this square-root dependence on the ratio of the Petermann factors holds for both Hermitian and symmetric $M$ if we additionally impose unitarity.

\subsubsection{When $M$ is a rank-one projector}
When $M$ is a projector, $M^\dagger = M$. Combining this with the condition that $M$ is a symmetric matrix, we have $M = M^T = M^* = M^\dagger$. Together with $M^2 = M$, we then have
\begin{align}
    \langle L|L\rangle = \langle R^*|M|R^*\rangle = \langle L|R^*\rangle = \langle R^*|L\rangle.
\end{align}
When the rank of $M$ is 1, we can write it as $M = |c\rangle \langle c|$ with a $\boldsymbol\lambda$-independent real vector with a unit length $|c\rangle$. Then, $|L\rangle = M|R^*\rangle$ implies that $|L\rangle = \langle c|R^*\rangle |c\rangle$. Taking its complex conjugate, $|L^*\rangle = \langle c|R^*\rangle^* |c\rangle = \frac{\langle c|R^*\rangle^*}{\langle c|R^*\rangle}|L\rangle$. This means that $|L\rangle$ and $|L^*\rangle$ are different by an overall phase factor, which possibly depends on $\boldsymbol\lambda$. Thus we can write $|L^*\rangle = e^{i \phi (\boldsymbol\lambda)}|L\rangle$. 
Then,
\begin{align}
    \langle L|L\rangle^2 &= \langle L|R^*\rangle\langle R^*|L\rangle
    =
    \langle L^*|R^*\rangle\langle R^*|L^*\rangle
    \notag \\
    &= \langle R|L\rangle\langle L|R\rangle
    = |\langle L|R\rangle|^2.
\end{align}
Therefore,
\begin{align}
    \frac{\langle R|R\rangle}{\langle L|L\rangle} 
    =
    \frac{\langle R|R\rangle\langle L|L\rangle}{\langle L|L\rangle^2}
    =
    \frac{\langle R|R\rangle \langle L|L\rangle}{|\langle L|R\rangle|^2} = K.
\end{align}
Thus, from Eq.~(\ref{eq:gafa2}), we obtain
\begin{align}
    A_g (\mathcal{C})
    &=
    \frac{K_T}{K_0},
\end{align}
which is the same as in the case of $|L\rangle = M|R\rangle$ with a Hermitian projector $M$. However, we note that the proof we presented here only works for rank-one projector, whereas the result in the case of $|L\rangle = M|R\rangle$ with a Hermitian projector $M$ holds for projectors with any rank.

An important example in which $M$ is both symmetric and a rank-1 projector is when $|L\rangle$ is a real vector independent of $\boldsymbol\lambda$.
As before, we can then choose $M = |L\rangle \langle L|$ with a normalized $|L\rangle$, and we see that $|L\rangle \propto M|R^*\rangle$.

\setcounter{subsection}{0} 
\renewcommand{\thesubsection}{\Alph{subsection}$^{\boldsymbol{\prime}}$}
\subsection{$|R\rangle = M|L\rangle$ with a Hermitian $M$}
The argument for $|R\rangle = M|L\rangle$ with a Hermitian $M$ 
runs parallel to that for $|L\rangle = M|R\rangle$. Writing down the final result, the adiabatic amplification factor is
\begin{align}
    A_g (\mathcal{C})
    &=
    \sqrt{K_T \frac{\langle L_T|M^2| L_T\rangle}{\langle L_T |L_T\rangle} \frac{1}{K_0}\frac{\langle L_0|L_0\rangle}{\langle L_0|M^2|L_0\rangle}}.
\end{align}
If we further impose that $M$ be a unitary matrix, we obtain
\begin{align}
    A_g (\mathcal{C})
    &=
    \sqrt{\frac{K_T}{K_0}}.
\end{align}

Instead, if we assume that $M$ is a projector, the relation
\begin{align}
    \langle R|R\rangle = \langle R|L\rangle = \langle L|R\rangle
\end{align}
holds and thus
\begin{align}
    \frac{\langle R|R\rangle}{\langle L|L\rangle} 
    =
    \frac{|\langle L|R\rangle|^2}{\langle R|R\rangle \langle L|L\rangle} = \frac{1}{K}.
\end{align}
This is the inverse of Eq.(\ref{eq:rrllk}). As a consequence, we obtain
\begin{align}
    A_g (\mathcal{C})
    &=
    1,
\end{align}
namely, there is no geometric amplification.
An important example where $M$ is both Hermitian and a projector is when $|R\rangle$ can be taken independent of $\boldsymbol\lambda$. When $|R\rangle$ is independent of $\boldsymbol\lambda$, $\boldsymbol{\mathcal{A}}^{LR} = \boldsymbol{\mathcal{A}}^{RR} = 0$ holds, and thus we can also directly see from Eq.~(\ref{eq:defgaf}) that the geometric amplification factor is 1.

\subsection{$|R\rangle = M|L^*\rangle$ with a symmetric $M$}
After calculations similar to the ones for $|L\rangle = M|R^*\rangle$, we obtain that the geometric amplification factor is
\begin{align}
    A_g (\mathcal{C})
    &=
    \sqrt{K_T \frac{\langle L_T^*|M^* M| L_T^*\rangle}{\langle L_T |L_T\rangle}\cdot \frac{1}{K_0}\frac{\langle L_0|L_0\rangle}{\langle L_0^*|M^* M|L_0^*\rangle}}.
\end{align}
If we further impose that $M$ be a unitary matrix, we obtain
\begin{align}
    A_g (\mathcal{C})
    &=
    \sqrt{\frac{K_T}{K_0}},
\end{align}
as before.

Instead, if we assume that $M$ is a rank-1 projector, the relation
\begin{align}
    \langle R|R\rangle^2 = |\langle L|R\rangle|^2
\end{align}
can be proved, and thus
\begin{align}
    \frac{\langle R|R\rangle}{\langle L|L\rangle} 
    =
    \frac{|\langle L|R\rangle|^2}{\langle R|R\rangle \langle L|L\rangle} = \frac{1}{K}.
\end{align}
As a consequence, we obtain
\begin{align}
    A_g (\mathcal{C})
    &=
    1,
\end{align}
namely, there is again no geometric amplification.
We note that, similar to the case of $|L\rangle = M|R^*\rangle$, when the symmetric matrix $M$ is also a projector, the derivation above works only when the rank of the projector is one.

\section{Summary of different cases}
\label{sec:sum}

We provide a summary of the previous section in Table~\ref{tab:tab}.
We also summarize here some representative Hamiltonians and systems in which each of the cases holds.

\begin{itemize}
\item Important examples of the case in which $|L\rangle = M|R^*\rangle$ or $|R\rangle = M|L^*\rangle$ with symmetric and unitary $M$ are when the Hamiltonian is related to a symmetric Hamiltonian via a unitary transformation. A special case includes when the Hamiltonian itself is a symmetric matrix. The geometric amplification factor of such systems is $A_g (\mathcal{C}) = \sqrt{\frac{K_T}{K_0}}$.
Various classical reciprocal systems, such as mechanical metamaterials~\cite{ghatak}, mass-spring systems~\cite{Nassar:2018}, and inductor-capacitor systems~\cite{Stegmaier:2024} fall into this category. We will encounter this situation as special cases of the two examples of Sec.~\ref{sec:models}, where one is a two-level system with generally non-reciprocal hoppings that in a special case can be transformed into reciprocal form, and the other is a model derived from a mechanical metamaterial~\cite{ghatak}.
\item Important examples of the case in which $|L\rangle = M|R\rangle$ or $|L\rangle = M|R^*\rangle$ with $M$ being a (rank-1) projector are systems in which $|L\rangle$ is independent of the parameter $\boldsymbol\lambda$. The geometric amplification factor of such systems is $A_g (\mathcal{C}) = \frac{K_T}{K_0}$. We will see in Sec.~\ref{sec:models} an example of such a situation using the model derived from a mechanical metamaterial~\cite{ghatak}.
\item Similarly, the case in which $|R\rangle = M|L\rangle$ or $|R\rangle = M|L^*\rangle$ with $M$ being a (rank-1) projector is obtained when $|R\rangle$ is independent of the parameter $\boldsymbol\lambda$. Such a situation should also be realizable using mechanical metamaterials of a similar setup. However, we note that in these special cases there is no geometric amplification, namely $A_g (\mathcal{C}) = 1$.
\end{itemize}

We note that, in principle, the relations such as $|L\rangle = M|R^*\rangle$ need to hold only for the particular state in which we are interested. Requiring, for example, the Hamiltonian to be symmetric results in $|L\rangle = |R^*\rangle$ for all the eigenstates, which is more strict than the condition to hold for a particular state.

\textbf{\begin{table*}[htp]
\begin{center}
\caption{Summary of the adiabatic amplification factor $A_g (\mathcal{C})$ when it only depends on the initial and final points of the adiabatic path in the parameter space}
\label{tab:tab}
\vspace{1em}
\begin{tabular}{l||c|c|c}
 & $M$ is Hermitian & $M$ is Hermitian and unitary& $M$ is a Hermitian projector \\
 \hline
 $|L\rangle = M|R\rangle$ & $A_g (\mathcal{C}) = \sqrt{\frac{K_T}{K_0} \frac{\langle R_T|R_T\rangle}{\langle R_T |M^2| R_T\rangle} \frac{\langle R_0|M^2|R_0\rangle}{\langle R_0|R_0\rangle}}$ & $A_g (\mathcal{C}) = \sqrt{\frac{K_T}{K_0}}$ & $A_g(\mathcal{C}) = \frac{K_T}{K_0}$\\ 
 \hline
  $|R\rangle = M|L\rangle$ & $A_g (\mathcal{C}) = \sqrt{\frac{K_T}{K_0} \frac{\langle L_T|M^2| L_T\rangle}{\langle L_T |L_T\rangle} \frac{\langle L_0|L_0\rangle}{\langle L_0|M^2|L_0\rangle}}$ & $A_g (\mathcal{C}) = \sqrt{\frac{K_T}{K_0}}$ & $A_g(\mathcal{C}) = 1$\\
  \hline
\end{tabular}
\\
\vspace{1em}
\begin{tabular}{l||c|c|c}
 & $M$ is symmetric & $M$ is symmetric and unitary & $M$ is a symmetric rank-1 projector \\
 \hline
  $|L\rangle = M|R^*\rangle$ & $A_g (\mathcal{C}) = \sqrt{\frac{K_T}{K_0} \frac{\langle R_T|R_T\rangle}{\langle R_T^* |M^* M| R_T^*\rangle} \frac{\langle R_0^*|M^* M|R_0^*\rangle}{\langle R_0|R_0\rangle}}$ & $A_g (\mathcal{C}) = \sqrt{\frac{K_T}{K_0}}$ & $A_g(\mathcal{C}) = \frac{K_T}{K_0}$\\
 \hline
 $|R\rangle = M|L^*\rangle$ & $A_g (\mathcal{C}) = \sqrt{\frac{K_T}{K_0} \frac{\langle L_T^*|M^* M| L_T^*\rangle}{\langle L_T |L_T\rangle} \frac{\langle L_0|L_0\rangle}{\langle L_0^*|M^* M|L_0^*\rangle}}$ & $A_g (\mathcal{C}) = \sqrt{\frac{K_T}{K_0}}$ & $A_g(\mathcal{C}) = 1$\\
 \hline
\end{tabular}
\end{center}
\end{table*}}

\renewcommand{\thesubsection}{\Alph{subsection}}
\section{Models}
\label{sec:models}
We now examine concrete models to show how the general framework we have developed can be applied.
We first consider a paradigmatic two-level non-Hermitian Hamiltonian, and we then discuss a model of a non-Hermitian one-dimensional robotic metamaterial.

\subsection{Two-level Hamiltonian}
We first consider a non-Hermitian two-level Hamiltonian
\begin{align}
    H(\Delta, J, \delta) = \begin{pmatrix} -\Delta & J + \delta \\ J - \delta & \Delta \end{pmatrix}, \label{eq:ham2lev}
\end{align}
which depends on three real-valued parameters $\boldsymbol\lambda = (\Delta, J, \delta)$.
Such non-Hermitian two-level Hamiltonians have been explored using coupled mechanical oscillators in Refs.~\cite{Anandwade2023,Singhal2023} to detect the non-Hermitian adiabatic Berry's phase.

The eigenvalues of this Hamiltonian are
\begin{align}
    E_\pm = \pm \sqrt{\Delta^2 + J^2 - \delta^2}.
\end{align}
Assuming that we only work in the regime $\Delta^2 + J^2 - \delta^2 \ge 0$, both eigenvalues are real and thus we can expect the adiabatic theorem to hold.

The corresponding non-normalized right and left eigenstates are
\begin{align}
    |R_\pm\rangle &= 
    \begin{pmatrix}
    -\Delta \pm \sqrt{\Delta^2 + J^2 - \delta^2} \\ J - \delta
    \end{pmatrix}, \\
    |L_\pm\rangle &= 
    \begin{pmatrix}
    -\Delta \pm \sqrt{\Delta^2 + J^2 - \delta^2} \\ J + \delta
    \end{pmatrix}.
\end{align}

The Petermann factor is identical for both ($+$ and $-$) eigenstates, and  is given by
\begin{align}
    K = \frac{\Delta^2 + J^2}{\Delta^2 + J^2 - \delta^2}.
\end{align}
The gauge-invariant Berry connection difference is
\begin{align}
    &\boldsymbol{\mathcal{A}}^{LR}_\pm - \boldsymbol{\mathcal{A}}^{RR}_\pm
    \equiv
    i\frac{\langle L_\pm|\nabla_{\boldsymbol\lambda} R_\pm\rangle}{\langle L_\pm|R_\pm\rangle}
    -
    i\frac{\langle R_\pm|\nabla_{\boldsymbol\lambda} R_\pm\rangle}{\langle L_\pm|R_\pm\rangle}
    \notag \\
    &=
    \frac{i \delta/2}{(\Delta^2 + J^2)E_\pm^2}
    \begin{pmatrix}
    \Delta \delta + JE_\pm,
    &
    J \delta - \Delta E_\pm,
    &
    -\Delta^2 - J^2
    \end{pmatrix}.
\end{align}
The Berry curvature difference is
\begin{align}
    \boldsymbol{\Omega}^{LR}_\pm - \boldsymbol{\Omega}^{RR}_\pm
    &\equiv
    \nabla_{\boldsymbol\lambda} \times \left( \boldsymbol{\mathcal{A}}^{LR}_\pm - \boldsymbol{\mathcal{A}}^{RR}_\pm \right)
    \notag \\
    &=
    \pm i \frac{(\Delta, J, \delta)}{2(\Delta^2 + J^2 - \delta^2)^{3/2}},
\end{align}
which is purely imaginary.

We can see that, when $\Delta = 0$, the $\Delta$-component of the Berry curvature is zero, which means that the Berry curvature vanishes in the $J$-$\delta$ plane. We will show below that, in this case, the adiabatic amplification is solely determined by the ratio of the Petermann factors of the initial and final points of the path in the parameter space. We will also examine, as an example of a case in which the adiabatic amplification factor does not only depend on the ratio of the Petermann factors, the adiabatic amplification  for fixed $J$ and $\delta$ when only $\Delta$ is varied.

\subsubsection{When $\Delta = 0$}

When $\Delta = 0$, the Hamiltonian 
\begin{align}
    H(0, J, \delta) = \begin{pmatrix} 0 & J + \delta \\ J - \delta & 0 \end{pmatrix}
\end{align}
is unitarily related to a reciprocal (symmetric) Hamiltonian. In fact, after a unitary transformation by the fixed matrix $U = \frac{1}{\sqrt{2}}\begin{pmatrix} 1 & -i \\ -i & 1\end{pmatrix}$, the Hamiltonian becomes 
\begin{align}
    UH(0, J, \delta)U^\dagger
    = \begin{pmatrix} i \delta  & J \\ J & -i\delta \end{pmatrix},
\end{align}
which is a symmetric non-Hermitian matrix.
According to the general discussion given in Sec.~\ref{sec:reciprocal}, the adiabatic amplification factor is then simply given by $\sqrt{K_T/K_0}$, where $K_T$ and $K_0$ are the Petermann factors at the final and initial points of the path in  parameter space.

\begin{figure}[t]
    \centering
\includegraphics[width=\linewidth]{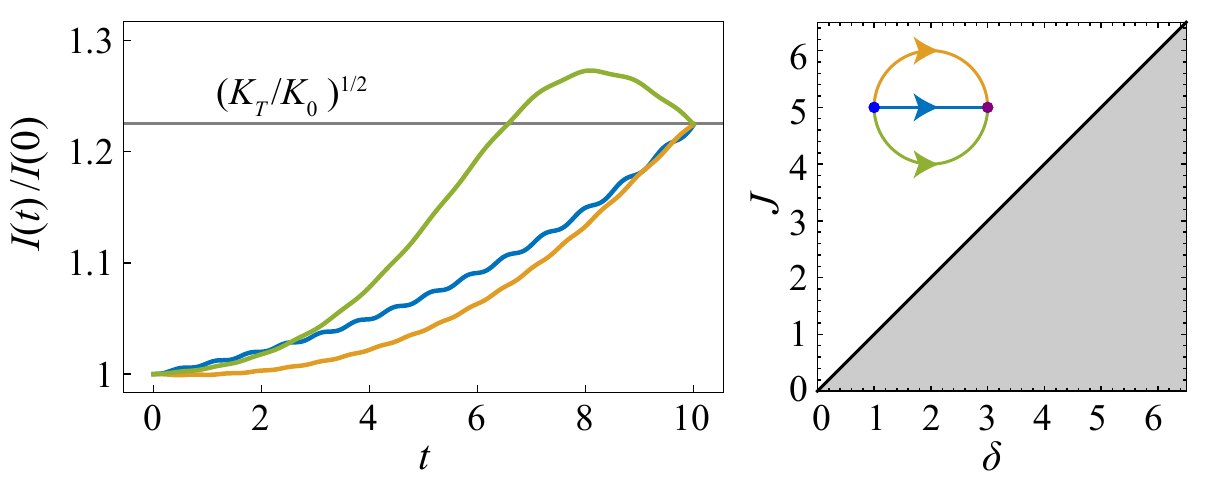}
    \caption{Numerical simulation of adiabatic amplification for the two-level Hamiltonian Eq.~(\ref{eq:ham2lev}) with $\Delta = 0$. (Left) Intensity amplification
    factor
    as a function of time for three different paths in the $J$-$\delta$ plane, traversed over a time duration $T=10$. The theoretically expected adiabatic amplification factor $\sqrt{K_T/K_0}$ is indicated by the horizontal line, to which all three curves converge. (Right) The three paths in the parameter space of $(\delta,  J)$. The shaded region in the right panel is the $PT$-broken region where the energy eigenvalues become complex. The three paths are all confined to the $PT$-symmetric region, where the energies are real. The colors of the three curves in the left panel match the colors of the paths described in the right panel.}
    \label{fig:twolevel}
\end{figure}

Let us numerically confirm this prediction by choosing several different paths in the $\delta$-$J$ plane.
In Fig.~\ref{fig:twolevel} we fix $\Delta = 0$ and we change $(\delta, J)$ from $(1,5)$ to $(3,5)$ along three different paths, which we traverse with a constant speed over a time interval $T=10$.
We see that, while the detailed time dependence of the norm of the state vector  during the adiabatic change does depend on the particular path one chooses, they all converge to the same value, which is simply $\sqrt{K_T/K_0}$. 
We note that small oscillations observed for the middle (blue) line of the left panel of Fig.~\ref{fig:twolevel} are due to the non-adiabaticity of the process; if we make the time interval $T$ larger, the oscillation goes away.

\subsubsection{When only $\Delta$ is varied}

Fixing $J$ and $\delta$ and varying only $\Delta$, one obtains adiabatic amplification in the one dimensional parameter space of $\Delta$. As we have noted in the general discussion, when the parameter space is one dimensional, the adiabatic amplification is determined solely by the properties at the initial and final points in the parameter space. Varying $\Delta$ from 
$\Delta_0$ to $\Delta_T$, 
the adiabatic amplification is given by the integral of the $\Delta$-component of the Berry connections as
\begin{align}
    A_g (\mathcal{C})
    &=
    \exp
    \left[ - 2\int_{\Delta_{0}}^{\Delta_{T}} \mathrm{Im}\left[ \mathcal{A}^{LR}_\pm (\Delta) - \mathcal{A}^{RR}_\pm (\Delta)\right] d\Delta \right]
    \notag \\
    &=
    \exp
    \left[ - \delta \int_{\Delta_{0}}^{\Delta_{T}} \frac{\Delta \delta \pm J\sqrt{\Delta^2 + J^2 - \delta^2}}{(\Delta^2 + J^2)(\Delta^2 + J^2 - \delta^2)}
    d\Delta \right],
\end{align}
where the upper (lower) sign corresponds to the adiabatic amplification of the $|R_+\rangle$ ($|R_-\rangle$) state.
This integral can be performed analytically. After simplifying the expression, one obtains that
\begin{align}
    A_g (\mathcal{C})
    =
    \frac{1 \mp \sqrt{1 - K_T/K^{(0)}}}{1 \mp \sqrt{1 - K_0/K^{(0)}}}, \label{eq:alongdelta}
\end{align}
where $K_0$ and $K_T$ are the Petermann factors at $\Delta_0$ and $\Delta_T$, respectively, and $K^{(0)} \equiv J^2/(J^2 - \delta ^2)$ is the Petermann factor at $\Delta = 0$.

One can see from Eq.~(\ref{eq:alongdelta}) that, unlike in the case of adiabatic amplification in the parameter space of $(\delta, J)$ with $\Delta = 0$, the expression is not just a function of $K_T / K_0$, but it has a different dependence on the Petermann factors at the initial and final points.

\subsection{One-dimensional robotic metamaterial}
To illustrate the effects of adiabatic amplification in a concrete physical setting, we adopt the model underlying  recent experiments on nonreciprocal robotic metamaterials~\cite{ghatak}. 
As detailed below, this model provides free and independent control of the profiles of a particular pair of right and left eigenstates, so that it can be used to realize the different cases mentioned above.

We will examine amplification properties of a zero mode, which is induced by the coupling topology (connectivity of the oscillators) and symmetry of the structure. As we show, making use of velocity-dependent friction, we can make such a zero mode the most stable mode of the system, suitable for adiabatic state control. 
Furthermore, it has been predicted that the behavior of the Petermann factor of this zero mode dictates a transition in the static behavior of the system, where it becomes directionally unstable \cite{Schomerus2020}. This instability occurs when the right and left eigenstates of the zero mode become localized at opposite ends of the structure. In the unstable phase, the structure then serves as a directional amplifier, which is sensitive to an input source at the end where the left eigenstate is localized, and produces an exponentially amplified output at the end where the right eigenstate is localized.
Furthermore, this strong response does not occur when the source and detector positions are interchanged. In reciprocal systems, the right and left states have exactly the same mode profile, so that this effect does not occur. 
Analysis of the Green's function then reveals that the Petermann factor serves as an order parameter for this transition, as it diverges in the limit of a large system when the system is directionally unstable, whereas it converges to a finite value when the system is stable \cite{Schomerus2020}.
The effect can be used for nonreciprocal sensing \cite{Budich2020}, and it also appears in open nonreciprocal quantum-optical systems \cite{Wanjura2020,McDonald2020}. Furthermore, we note that while often occurring along with it, this effect is distinct from the non-Hermitian skin effect, where the latter  pertains to a high sensitivity of the spectrum to the boundary conditions, which can also occur in reciprocal systems  \cite{Okuma2020,Hofmann2020}. 
Based on the theory developed in the present work, we establish the dynamical signatures of this instability as the system is steered adiabatically across the phase transition. We also contrast this situation with the reciprocal setting in which the transition does not exist.

\subsubsection{Model details}
The metamaterial is constructed out of $N$ mechanical oscillators whose couplings are mediated by electronic components, whereby the reciprocity conditions of purely mechanical systems can be circumvented. Spectral and topological notions can be applied to the physics of this system when it operates in the linear regime.
The equation of motion
\begin{equation}
    \frac{d^2}{dt^2}\mathbf{x}=-\mathcal{M}\mathbf{x}
    \label{eq:eom}
\end{equation}
of the oscillator amplitudes $\mathbf{x}$ is then governed by a generally nonsymmetric dynamical matrix $\mathcal{M}\neq \mathcal{M}^T$, whose spectral properties form the basis of our considerations. In such structures zero frequency modes appear whenever there is an imbalance between the number of degrees of freedom and the number of constraints in a suitably chosen coupling topology, which is also encoded in $\mathcal{M}$ \cite{Kane:2014,Chen:2014,Huber:2016,ghatak}.

The equation of motion Eq.~(\ref{eq:eom}) is a second order differential equation. To apply the theory of adiabatic amplification developed in this paper, we  rewrite the equation of motion in terms of  first order differential equations, i.e., in the form of the Schr\"odinger equation, following the steps in Ref.~\cite{Kane:2014} where this analogy was pursued to illuminate the origin of the zero modes. This can be achieved by exploiting that the dynamical matrix can be decomposed as $\mathcal{M}=\mathcal{Q}\mathcal{R}$, where $\mathcal{Q}^T$ and $\mathcal{R}$ are $(N^\prime \times N)$-dimensional matrices with $N^\prime \le N$ and the ranks of $\mathcal{Q}^T$ and $\mathcal{R}$ are both $N^\prime$.
We can then rewrite the equation of motion, Eq.~(\ref{eq:eom}), in terms of  first-order differential equations by introducing  effective momentum degrees of freedom $\mathbf{p}$, whose number of elements $N^\prime$ may differ from $N$, as
\begin{equation}
    i\frac{d}{dt}\left(\begin{array}{c}
    \mathbf{x}\\
    \mathbf{p}
    \end{array}\right)=
i\left(\begin{array}{cc}
    0&  \mathcal{Q}\\
   - \mathcal{R} &  0
\end{array}\right)\left(\begin{array}{c}
    \mathbf{x}\\
    \mathbf{p}
    \end{array}\right).
    \label{eq:eom2}
\end{equation}
For the precise definition of $\mathbf{p}$, see~\footnote{Since $\mathcal{Q}$ is $N \times N^\prime$ and the rank of $\mathcal{Q}$ is $N^\prime$, there exists an $N^\prime \times N$ left-inverse matrix of $\mathcal{Q}$, which we denote by $\mathcal{Q}_\mathrm{left}^{-1}$, satisfying $\mathcal{Q}_\mathrm{left}^{-1} \mathcal{Q} = \mathbb{I}_{N^\prime}$. Here, $\mathbb{I}_{N^\prime}$ is an $N^\prime \times N^\prime$ identity matrix. Using this left-inverse matrix, we define $\mathbf{p}$ by $\mathbf{p} \equiv \mathcal{Q}_\mathrm{left}^{-1} \frac{d}{dt}\mathbf{x}$. It is then straightforward to show that the first-order differential equation~(\ref{eq:eom2}) is satisfied.}.
This equation of motion takes the form of a Schr\"odinger equation when we identify the effective Hamiltonian and the state vector as
\begin{align}
H &= i\begin{pmatrix}
    0&  \mathcal{Q}\\
    -\mathcal{R} &  0
\end{pmatrix},&
|\psi\rangle &= 
\begin{pmatrix}
    \mathbf{x}\\
    \mathbf{p}
    \end{pmatrix}.
    \label{eq:effHamiltonian}
\end{align}
The effective Hamiltonian is non-Hermitian if $\mathcal{Q}^\dagger\neq \mathcal{R}$ and effectively nonreciprocal (nonsymmetric) if $\mathcal{Q}^T\neq\mathcal{R}$ \footnote{For $\mathcal{Q}^T=\mathcal{R}$ the effective Hamiltonian can be brought into  symmetric form by a unitary transformation $UHU^{-1}$, where $U=\mathrm{diag}(\mathbb{I}_{N},i\mathbb{I}_{N'})$. This condition matches the definition of reciprocity in ordinary mechanics, where $\mathcal{Q}$ and $\mathcal{R}$ are furthermore real. The mathematical alternative $\mathcal{Q}^T=-\mathcal{R}$ results, for real $Q$ and $R$, in a negative semi-definite matrix $M$, which corresponds to a mechanically unstable system.} .

In general, Hamiltonians of this structure have $N-N'$ zero modes.
To see this, note that since $\mathcal{R}$ is $N^\prime \times N$ and its rank is $N^\prime$, according to the rank-nullity theorem there must then be $N-N^\prime$ linearly independent $N$-component vectors $\mathbf{x}_i$, which satisfy $\mathcal{R}\mathbf{x}_i = \mathbf{0}$. Then, the vectors $\displaystyle \begin{pmatrix} \mathbf{x}_i \\ \mathbf{0}_{N^\prime}\end{pmatrix}$, with $\mathbf{0}_{N^\prime}$ being an $N^\prime$-component zero vector, are right eigenstates of $H$ with zero eigenvalue. The left eigenstates can be similarly constructed as follows.
Since $\mathcal{Q}^\dagger$ is $N^\prime \times N$ with rank $N^\prime$, there again must be $N-N^\prime$ linearly independent $N$-component vectors $\tilde{\mathbf{x}}_i$ that satisfy $\mathcal{Q}^\dagger \tilde{\mathbf{x}}_i = \mathbf{0}$. Then, the vectors $\displaystyle \begin{pmatrix} \tilde{\mathbf{x}}_i \\ \mathbf{0}_{N^\prime}\end{pmatrix}$ are the right eigenstates of $H^\dagger$, 
so that the adjoint of these vectors gives the left eigenstates of $H$.
An important feature to notice here is that the right and left eigenstates of $H$ are completely determined by $\mathcal{R}$ and $\mathcal{Q}$, respectively. We also note that the effective momentum components of the zero modes are all zero for both the right and left eigenstates.

\subsubsection{Coupling configuration and mode profiles}

Here, we focus on the case $N'=N-1$, where there exists a single zero mode. Specifically, we refer to the experiment \cite{ghatak}, which was  designed so that the effective Hamiltonian in Eq.~\eqref{eq:eom2} corresponds to a nonreciprocal Su-Schrieffer-Heeger (SSH) chain. This is achieved by setting
\begin{equation}\mathcal{R}_{nm}=-a\delta_{nm}+b\delta_{n,m-1},\quad\mathcal{Q}_{nm}=-a'\delta_{nm}+b'\delta_{n-1,m}
,
\end{equation}
which, in the setting of the effective Hamiltonian \eqref{eq:effHamiltonian}, corresponds to alternating directed couplings $a,b,a,b,\ldots$ as one moves along a one dimensional chain of $2N-1$ sites in one direction, and analogously with $a'$, $b'$ for the other direction.
This coupling configuration  places the zero mode into the middle of a band gap of size
\begin{equation}
    \Delta=2\sqrt{(|a|-|b|)(|a'|-|b'|)}.
    \label{eq:bandgap}
\end{equation}

The position amplitudes of the right and left eigenstates of the zero mode are
given by
\begin{equation}
    \langle n|R \rangle = c_R \left(\frac{a}{b}\right)^n,\quad
    \langle L|n \rangle = c_L \left(\frac{a'}{b'}\right)^n.
\label{eq:modeprofiles}
\end{equation}
As we noted earlier, the momentum amplitudes of the right and left eigenstates all vanish for the zero mode.
This gives rise to exponentially localized mode profiles, whose confinement to the near end (around $n=1$) or far end (around $n=N$) of the system can be determined from the indices
\begin{equation}
   \xi= \mathrm{sgn}\,(|b|^2-|a|^2)
   \label{eq:xi}
\end{equation}
for the right eigenstate, 
and 
\begin{equation}
   \eta= \mathrm{sgn}\,(|b'|^2-|a'|^2)
\end{equation}
for the left eigenstate. As we will see below, the relevant difference in the mode profiles depends only on the nonreciprocity parameter
\begin{equation}
    \varepsilon=\frac{ab'-ba'}{ab'+ba'}. \label{eq:nonrec}
\end{equation}

\begin{figure*}[t]
  \centering
\includegraphics[width=0.8\linewidth]{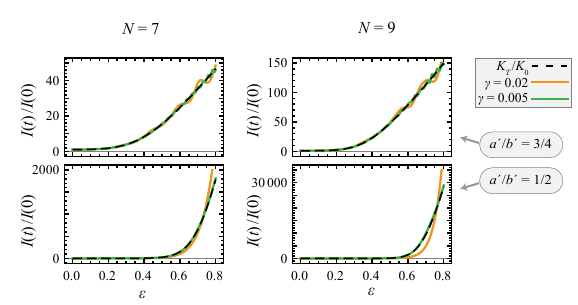}
\caption{Comparison between intensity dynamics $I_T/I_0$ (colored curves) and ratio of Petermann factors $K_T/K_0$  (dashed black curves) in a robotic metamaterial in which the left eigenstate of a zero mode is fixed (with two different decay factors $a'/b'$), while the right eigenstate evolves adiabatically as the nonreciprocity parameter $\varepsilon$ is changed at a constant rate $\gamma$, see Eq.~\eqref{eq:varept}. The intensity dynamics approach the prediction of our theory, according to which the amplification factor is given by the ratio of Petermann factors, when the rate is decreased.
The Petermann factor becomes large when the right eigenstate relocalizes to the other side to the system, so that its overlap with the left state becomes small. This would become a sharp transition for large systems (here we have $N=7$ or 9 mechanical oscillators). 
}\label{fig1}
\end{figure*}

We see from its definition, Eq.~(\ref{eq:petermann}), that the Petermann factor directly captures differences of the mode profiles. In particular, the Petermann factor becomes large when the overlap $|\langle L|R\rangle|^2$  is small, which occurs when the two types of states are localized at opposite ends.
An explicit expression is found by inserting the mode profiles \eqref{eq:modeprofiles} into the definition \eqref{eq:petermann} \cite{Schomerus2020}.
In the limit of a large system $N\gg 1$, the Petermann factor then behaves as 
\begin{equation}
K\sim \left\{
  \begin{array}{ll}
    K_\infty , & \xi\eta=1\,\hbox{\mbox{(stable)};} \\[.2cm]
   \displaystyle 
   |K_\infty| g^N, 
   & \xi\eta=-1\,\hbox{\mbox{(unstable)}.}
  \end{array}
\right.
\end{equation}
where 
\begin{equation}
K_\infty= \frac{|aa'-bb'|^2}{(|a|^2-|b|^2)(|a'|^2-|b'|^2)}, 
\end{equation}
and 
$g=\min\left(|b/a|^{2\xi},|b'/a'|^{2\eta}\right)$.
Therefore, the Petermann factor either converges to a finite value, or increases exponentially with system size.
As indicated, this discriminates between two phases in which the system is stable or unstable against external dynamical perturbations, and in particular, driving by external forces at one of its edges, with the response exponentially amplified at the other edge. 

\subsubsection{Adiabatic amplification with a fixed left eigenstate}

We first consider the adiabatic state amplification for a fixed left eigenstate, hence, for fixed $a'$ and $b'$. This situation corresponds to $|L\rangle = M|R\rangle$ with a Hermitian projector $M$, and thus according to Table.~\eqref{tab:tab} the amplification factor for an adiabatic parameter change is given by the ratio of Petermann factors $K_T/K_0$.
Any deviations from this signals a departure from the adiabatic conditions. 

This ratio generally only depends on the nonreciprocity parameter $\varepsilon$
defined in Eq.~(\ref{eq:nonrec}). For instance, we can rewrite the Petermann factor in the stable phase of an infinite system as
\begin{equation}
K_\infty= 
    \displaystyle \frac{|(a'/b')^2-(1-\varepsilon)/(1+\varepsilon)|^2}{[
    |a'/b'|^2-|(1-\varepsilon)/(1+\varepsilon)|^2]
   (|a'/b'|^2-1)}.
\end{equation}
However, we note that while the zero-mode profiles only depend on the ratios $a/b$ and $a'/b'$, the adiabaticity condition also depends on the size \eqref{eq:bandgap} of the band gap $\Delta$ (as the relevant dimensionless time scale is $\Delta\,t$) and the system size $N$ (as this determines the sharpness of the instability transition and the size of the Petermann factor in the unstable phase).

To illustrate these points, we set $a=a'(1+\varepsilon)$, $b=b'(1-\varepsilon)$, and we increase the nonreciprocity parameter $\varepsilon$ at a constant rate
\begin{align}
    \varepsilon (t) = \gamma \Delta_0 t, \label{eq:varept}
\end{align}
from $\varepsilon_\mathrm{min}=0$ to $\varepsilon_\mathrm{max}=0.8$, where $\Delta_0=2\left||a'|-|b'|\right|$ is the band gap in the reciprocal system ($\varepsilon=0$), and $\gamma$ gives control of the degree of adiabaticity.

We performed numerical simulations of the time evolution of Eq.~(\ref{eq:eom2}) changing $\varepsilon(t)$ as Eq.~(\ref{eq:varept}), starting from a zero-mode right eigenstate of the Hamiltonian at $t = 0$. We choose two different values $a^\prime / b^\prime = 1/2$ and 3/4, and we show the intensity ratio $I(t)/I(0)$ as a function of time in Fig.~\ref{fig1}.
The system is in the stable region ($\xi \eta = 1$) at $t = 0$ and enters into the unstable region ($\xi \eta = -1$) at $\varepsilon = \frac{1 - |a^\prime/b^\prime|}{1 + |a^\prime / b^\prime|}$, which are 1/3 and 1/7 for $a^\prime / b^\prime = 1/2$ and 3/4, respectively. We note that the zero mode is more strongly localized for $a^\prime / b^\prime = 1/2$, resulting in larger amplification of the state after time evolution.

As seen from the figure, the zero-mode amplification becomes adiabatic when the rate $\gamma$ is sufficiently small, and the amplification factor $I(T)/I(0)\sim K_T/K_0$ then closely follows the prediction from our theory. 
Furthermore, this adiabatic regime is attained more easily (i.e., for relatively larger rates) when the system size $N$ is small, or the zero mode is less strongly confined ($|a/b|$ closer to 1). The directional instability transition is clearly visible as a drastic increase in the amplification factor, which reaches up to  $I(T)/I(0)\sim 30\,000$ under the displayed adiabatic conditions, and it exceeds this under nonadiabatic conditions.

\subsubsection{Adiabatic amplification for reciprocal Hamiltonian}

In the considered model, reciprocal systems are realized by setting $a=a'$ and $b=b'$. If these parameters are all real, the effective Hamiltonian is Hermitian and the time evolution is unitary, so that there is no amplification, even under non-adiabatic conditions. We therefore consider the setting where $a=a'$ can be complex and time-dependent, while we keep $b=b^\prime$ real and fixed; the latter constraint can always be achieved under static conditions by a suitable local gauge transformation.
When the Hamiltonian is reciprocal, $|R\rangle = |L^*\rangle$, and thus from Table.~\eqref{tab:tab}, the adiabatic amplification factor is given by $\sqrt{K_T/K_0}$.

\begin{figure*}[ht]
  \centering
\includegraphics[width=0.9\linewidth]{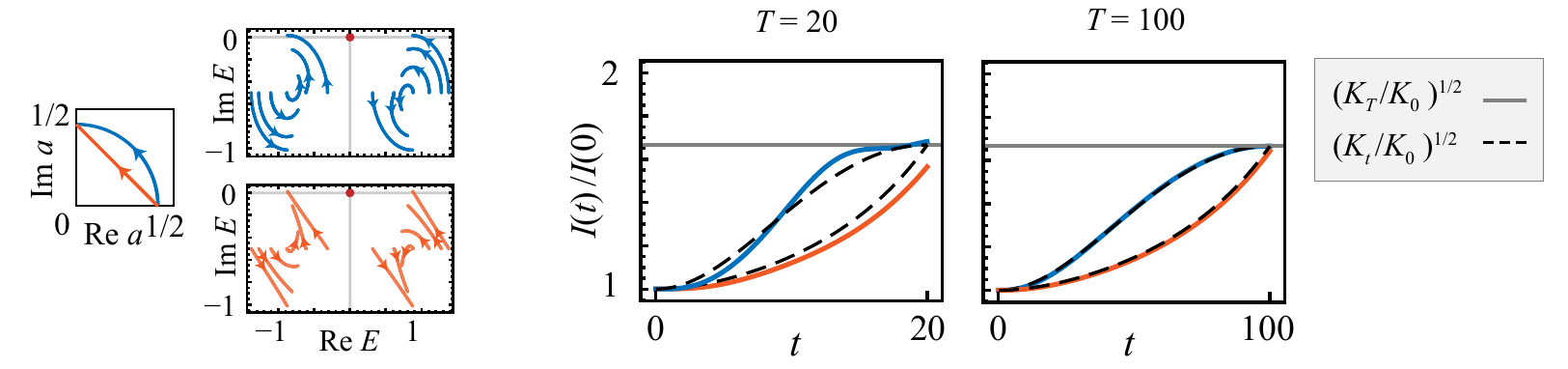}
\caption{Large panels: Comparison between intensity dynamics $I(T)/I(0)$ (colored curves) and $\sqrt{K_T/K_0}$  (dashed black curves) in a damped reciprocal robotic metamaterial ($N=7$ oscillators), with additional damping at a damping rate $\Gamma=1$. The parameters $b=b'=1$ are kept fixed
while the parameter  $a=a'$ is changed over a time $T=20$ (left) or $T=100$ (right) from $a_0=1/2$ to $a_T=i/2$, where the path is straight (orange curves) or on a circular arc (blue curves).
Side panels: Trajectories in the parameter space of complex $a$ and evolution of the complex energies while $a$ is changed along the trajectories; the zero mode energy is marked in red.
The Petermann factor increases as the system departs from Hermiticity. The intensity dynamics closely follows this behavior in the adiabatic regime, even though at the very end of the trajectory some of the eigenenergies enter into the upper half of the complex plane.
}\label{fig2}
\end{figure*}

\begin{figure*}[ht]
  \centering
\includegraphics[width=0.9\linewidth]{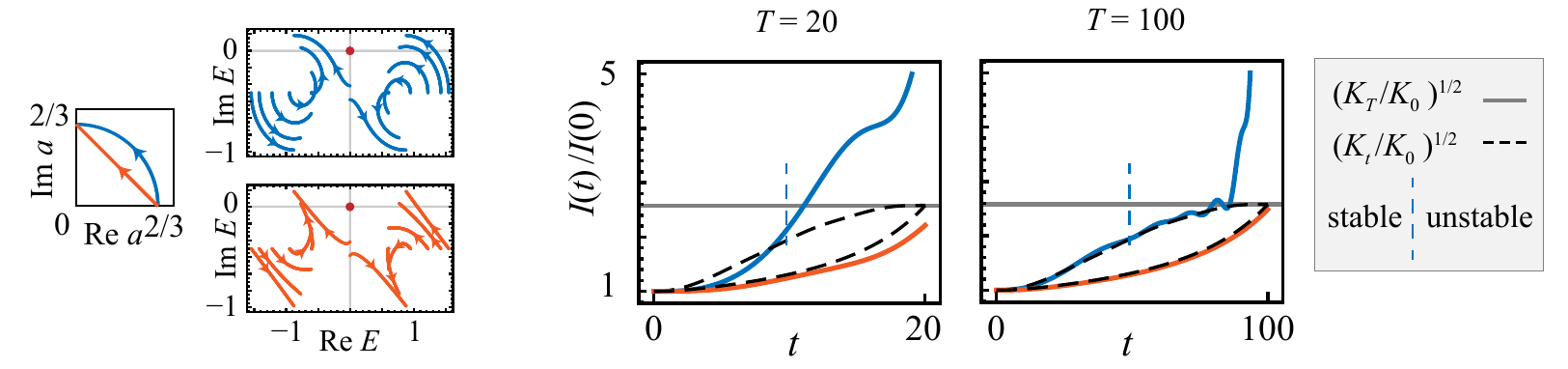}
\caption{As Fig.~\ref{fig2}, but for paths from $a_0=2/3$ to $a_T=2i/3$. As long as adiabatic conditions are observed, there is again good agreement between the numerical amplification factor and the prediction from the Petermann factor. For the circular path the adiabatic conditions now noticeably break down after the instantaneous eigenenergy of a bulk state moves into the upper half of the complex plane (at $t=0.49\,T$, indicated by the dashed vertical line), but for adiabatic conditions ($T=100$) this effect occurs with a significant delay. For the straight path, some eigenenergies enter into the upper half of the complex plane as well, but the intensity dynamics more robustly follows the prediction from the adiabatic theorem.
}\label{fig2b}
\end{figure*}

For this non-Hermitian reciprocal setting with complex parameters, the bulk band structure of the periodic system is complex as well, and in the finite system the zero mode will sit inside a cloud of complex resonance  eigenenergies that arise from the size-quantization of the bulk modes. Within the original definition of the system, our assumptions for adiabatic time evolution of the zero mode are therefore violated. However, the zero mode energy and  profiles survive the addition of any form of velocity-dependent friction, which therefore can be used to stabilize the adiabatic evolution of this state. We therefore modify the equations of motion    \eqref{eq:eom2}
to read
\begin{equation}
    i\frac{d}{dt}\left(\begin{array}{c}
    \mathbf{x}\\
    \mathbf{p}
    \end{array}\right)=
i\left(\begin{array}{cc}
    0&  \mathcal{Q}\\
   - \mathcal{R} &  -\Gamma \mathbb{I}
\end{array}\right)\left(\begin{array}{c}
    \mathbf{x}\\
    \mathbf{p}
    \end{array}\right),
    \label{eq:eom3}
\end{equation}
where $\Gamma$ quantifies the damping rate. This parameter then has the effect of moving the complex bulk resonance energies down into the complex plane as it is increased. We note that the introduction of $\Gamma$ does not break the reciprocity of the Hamiltonian.

Figures~\ref{fig2} and ~\ref{fig2b} show results for systems with fixed $b=b'\equiv b_0$, $\Gamma=b_0$, $N=7$, in which $a=a'$ is changed along two paths from a real initial position $a=a'=a_0$, to an imaginary final position  $a=a'=i a_0$. The path 
may be circular, $a=a'=a_0\exp(i\varphi)$ with $\varphi$ changing linearly in time from 0 to $\pi/2$, or straight. In Fig.~\ref{fig2} we set $a_0/b_0 =1/2$, which produces more strongly localized zero modes than for the choice $a_0/b_0=2/3$ shown in Fig.~\ref{fig2b}. For each setting we then compare the results for the dynamics of traversing these paths over a given time interval $T$. 

In all cases, the amplification factor agrees well with the prediction
$I(t)/I(0)=\sqrt{K_t/K_0}$ as long as the adiabaticity conditions are fulfilled.
We note that at the end of the paths, some bulk modes acquire a positive imaginary part in their energies, so that our assumptions of adiabatic evolution are violated. However, this has a noticeable effect only if the corresponding instantaneous states acquire a sufficient overlap with the evolving state, as seen for the circular path with  $a_0/b_0=2/3$. 
In particular, for $a_0/b_0=1/2$, where effective adiabaticity holds until the very end of both paths, the final amplification factor $I(T)/I(0)=\sqrt{K_T/K_0}$ agrees for both path shapes. 

Overall, the amplification factors are much smaller than in the nonreciprocal case with a fixed left eigenstate, which reflects the absence of directional instability in reciprocal systems.

\section{Conclusion}
\label{sec:conclusions}
We have studied the geometric contribution to the adiabatic amplification in non-Hermitian Hamiltonians, and we elucidated conditions under which the adiabatic amplification factor is determined solely by the initial and final points in the parameter space, and therefore does not depend on specific paths chosen to connect the two points. Our method provides an experimentally robust method to amplify the signal (state vector) to a desired intensity, which  depends neither on the path connecting the initial and final points nor on the speed of the change of the parameters along the chosen path, as long as  adiabatic conditions are observed. 
We have also identified conditions under which the adiabatic amplification factor can be written solely in terms of the ratio of Petermann factors of the states at the initial and final point. Our method thus provides a practical method to directly determine the Petermann factor, which generally quantifies the sensitivity of non-Hermitian systems to static and dynamical perturbations, but so far has been challenging to isolate in observable data. 

Generally, our findings open the door to reliably steer non-Hermitian systems into states with favorable properties, as is desirable for sensing and amplification. While we focused on these effects in classical systems, this could be extended to the quantum limited operation of devices in which analogous effective non-Hermitian descriptions arise, e.g., from a mean-field treatment in which quantum noise can then be incorporated via fluctuation-dissipation theorems \cite{Beenakker:1998}.

We note that we have mainly focused on the case in which the parameter space is simply connected. When the parameter space is not simply connected, such as a torus, the geometric amplification factor can depend on the path even when the condition Eq.~(\ref{eq:cond3}) is met. A particularly interesting situation is when two different paths $\mathcal{C}$ and $\mathcal{C}^\prime$ enclose singularities in the parameter space, such as an exceptional point. In such situations, the difference in the geometric amplification factors for the two paths should be topological, depending only on the topology of the closed path $\mathcal{C} - \mathcal{C}^\prime$. We should, however, note that when encircling an exceptional point, adiabaticity would be violated and thus the geometric amplification factor would not simply appear as the change of the norm. We leave it for the future to further investigate the property of the geometric amplification factor on such non-simply connected surfaces.

From a broader perspective, our results raise the prospect of identifying a wider range of dynamical effects that reveal the geometric properties of non-Hermitian physics. This could encompass non-adiabatic situations, generalizing the Aharonov-Anandan phase~\cite{Aharonov1987} known in Hermitian settings.
Non-adiabatic situations unique to non-Hermitian settings include sudden quenches across a non-Hermitian phase transition and systems stabilized at non-Hermitian eigenvalue degeneracies (exceptional points), which occur, for instance, at the phase transition of spontaneous PT symmetry breaking.
In analogy to the situation in Hermitian systems, the latter setting promises to reveal non-Abelian geometric effects in the driven dynamics, which would offer additional routes to manipulate the state of a system. 

\begin{acknowledgments}
We would like to thank Masahiko G. Yamada and one of the anonymous referees for pointing out issues related to non-simply connected parameter spaces.
H.S. wishes to thank AIMR, Tohoku University, where this work was carried out, for the kind hospitality during his stay.
T.O. is supported by JSPS KAKENHI Grant No. JP24K00548, JST PRESTO Grant No. JPMJPR2353, and JST CREST Grant No. JPMJCR19T1. This work has been initiated and supported by the Global Intellectual Incubation and Integration Laboratory (GI$^3$ Lab) Program provided by WPI-AIMR, Tohoku University.
\end{acknowledgments}

\bibliography{refs}

\begin{thebibliography}{66}%
\makeatletter
\providecommand \@ifxundefined [1]{%
 \@ifx{#1\undefined}
}%
\providecommand \@ifnum [1]{%
 \ifnum #1\expandafter \@firstoftwo
 \else \expandafter \@secondoftwo
 \fi
}%
\providecommand \@ifx [1]{%
 \ifx #1\expandafter \@firstoftwo
 \else \expandafter \@secondoftwo
 \fi
}%
\providecommand \natexlab [1]{#1}%
\providecommand \enquote  [1]{``#1''}%
\providecommand \bibnamefont  [1]{#1}%
\providecommand \bibfnamefont [1]{#1}%
\providecommand \citenamefont [1]{#1}%
\providecommand \href@noop [0]{\@secondoftwo}%
\providecommand \href [0]{\begingroup \@sanitize@url \@href}%
\providecommand \@href[1]{\@@startlink{#1}\@@href}%
\providecommand \@@href[1]{\endgroup#1\@@endlink}%
\providecommand \@sanitize@url [0]{\catcode `\\12\catcode `\$12\catcode `\&12\catcode `\#12\catcode `\^12\catcode `\_12\catcode `\%12\relax}%
\providecommand \@@startlink[1]{}%
\providecommand \@@endlink[0]{}%
\providecommand \url  [0]{\begingroup\@sanitize@url \@url }%
\providecommand \@url [1]{\endgroup\@href {#1}{\urlprefix }}%
\providecommand \urlprefix  [0]{URL }%
\providecommand \Eprint [0]{\href }%
\providecommand \doibase [0]{https://doi.org/}%
\providecommand \selectlanguage [0]{\@gobble}%
\providecommand \bibinfo  [0]{\@secondoftwo}%
\providecommand \bibfield  [0]{\@secondoftwo}%
\providecommand \translation [1]{[#1]}%
\providecommand \BibitemOpen [0]{}%
\providecommand \bibitemStop [0]{}%
\providecommand \bibitemNoStop [0]{.\EOS\space}%
\providecommand \EOS [0]{\spacefactor3000\relax}%
\providecommand \BibitemShut  [1]{\csname bibitem#1\endcsname}%
\let\auto@bib@innerbib\@empty
\bibitem [{\citenamefont {Bender}(2007)}]{Bender2007}%
  \BibitemOpen
  \bibfield  {author} {\bibinfo {author} {\bibfnamefont {C.~M.}\ \bibnamefont {Bender}},\ }\bibfield  {title} {\bibinfo {title} {{Making sense of non-Hermitian Hamiltonians}},\ }\href {https://iopscience.iop.org/article/10.1088/0034-4885/70/6/R03} {\bibfield  {journal} {\bibinfo  {journal} {Rep. Prog. Phys.}\ }\textbf {\bibinfo {volume} {70}},\ \bibinfo {pages} {947} (\bibinfo {year} {2007})}\BibitemShut {NoStop}%
\bibitem [{\citenamefont {Ashida}\ \emph {et~al.}(2020)\citenamefont {Ashida}, \citenamefont {Gong},\ and\ \citenamefont {Ueda}}]{Ashida2020}%
  \BibitemOpen
  \bibfield  {author} {\bibinfo {author} {\bibfnamefont {Y.}~\bibnamefont {Ashida}}, \bibinfo {author} {\bibfnamefont {Z.}~\bibnamefont {Gong}},\ and\ \bibinfo {author} {\bibfnamefont {M.}~\bibnamefont {Ueda}},\ }\bibfield  {title} {\bibinfo {title} {{Non-{H}ermitian physics}},\ }\href {https://www.tandfonline.com/doi/full/10.1080/00018732.2021.1876991} {\bibfield  {journal} {\bibinfo  {journal} {Adv. Phys.}\ }\textbf {\bibinfo {volume} {69}},\ \bibinfo {pages} {249} (\bibinfo {year} {2020})}\BibitemShut {NoStop}%
\bibitem [{\citenamefont {Gamow}(1928)}]{Gamow1928}%
  \BibitemOpen
  \bibfield  {author} {\bibinfo {author} {\bibfnamefont {G.}~\bibnamefont {Gamow}},\ }\bibfield  {title} {\bibinfo {title} {{Zur Quantentheorie des Atomkernes}},\ }\href {https://link.springer.com/article/10.1007/BF01343196} {\bibfield  {journal} {\bibinfo  {journal} {Z. Phys.}\ }\textbf {\bibinfo {volume} {51}},\ \bibinfo {pages} {204} (\bibinfo {year} {1928})}\BibitemShut {NoStop}%
\bibitem [{\citenamefont {Feshbach}(1958)}]{Feshbach1958}%
  \BibitemOpen
  \bibfield  {author} {\bibinfo {author} {\bibfnamefont {H.}~\bibnamefont {Feshbach}},\ }\bibfield  {title} {\bibinfo {title} {Unified theory of nuclear reactions},\ }\href {https://www.sciencedirect.com/science/article/abs/pii/0003491658900071} {\bibfield  {journal} {\bibinfo  {journal} {Ann. Phys.}\ }\textbf {\bibinfo {volume} {5}},\ \bibinfo {pages} {357} (\bibinfo {year} {1958})}\BibitemShut {NoStop}%
\bibitem [{\citenamefont {Hatano}\ and\ \citenamefont {Nelson}(1996)}]{Hatano1996}%
  \BibitemOpen
  \bibfield  {author} {\bibinfo {author} {\bibfnamefont {N.}~\bibnamefont {Hatano}}\ and\ \bibinfo {author} {\bibfnamefont {D.~R.}\ \bibnamefont {Nelson}},\ }\bibfield  {title} {\bibinfo {title} {{Localization Transitions in Non-Hermitian Quantum Mechanics}},\ }\href {https://doi.org/10.1103/PhysRevLett.77.570} {\bibfield  {journal} {\bibinfo  {journal} {Phys. Rev. Lett.}\ }\textbf {\bibinfo {volume} {77}},\ \bibinfo {pages} {570} (\bibinfo {year} {1996})}\BibitemShut {NoStop}%
\bibitem [{\citenamefont {Hatano}\ and\ \citenamefont {Nelson}(1997)}]{Hatano1997}%
  \BibitemOpen
  \bibfield  {author} {\bibinfo {author} {\bibfnamefont {N.}~\bibnamefont {Hatano}}\ and\ \bibinfo {author} {\bibfnamefont {D.~R.}\ \bibnamefont {Nelson}},\ }\bibfield  {title} {\bibinfo {title} {{Vortex pinning and non-Hermitian quantum mechanics}},\ }\href {https://doi.org/10.1103/PhysRevB.56.8651} {\bibfield  {journal} {\bibinfo  {journal} {Phys. Rev. B}\ }\textbf {\bibinfo {volume} {56}},\ \bibinfo {pages} {8651} (\bibinfo {year} {1997})}\BibitemShut {NoStop}%
\bibitem [{\citenamefont {Bender}\ and\ \citenamefont {Boettcher}(1998)}]{Bender1998}%
  \BibitemOpen
  \bibfield  {author} {\bibinfo {author} {\bibfnamefont {C.~M.}\ \bibnamefont {Bender}}\ and\ \bibinfo {author} {\bibfnamefont {S.}~\bibnamefont {Boettcher}},\ }\bibfield  {title} {\bibinfo {title} {{Real Spectra in Non-Hermitian Hamiltonians Having $PT$ Symmetry}},\ }\href {https://doi.org/10.1103/PhysRevLett.80.5243} {\bibfield  {journal} {\bibinfo  {journal} {Phys. Rev. Lett.}\ }\textbf {\bibinfo {volume} {80}},\ \bibinfo {pages} {5243} (\bibinfo {year} {1998})}\BibitemShut {NoStop}%
\bibitem [{\citenamefont {Makris}\ \emph {et~al.}(2008)\citenamefont {Makris}, \citenamefont {El-Ganainy}, \citenamefont {Christodoulides},\ and\ \citenamefont {Musslimani}}]{Makris2008}%
  \BibitemOpen
  \bibfield  {author} {\bibinfo {author} {\bibfnamefont {K.~G.}\ \bibnamefont {Makris}}, \bibinfo {author} {\bibfnamefont {R.}~\bibnamefont {El-Ganainy}}, \bibinfo {author} {\bibfnamefont {D.~N.}\ \bibnamefont {Christodoulides}},\ and\ \bibinfo {author} {\bibfnamefont {Z.~H.}\ \bibnamefont {Musslimani}},\ }\bibfield  {title} {\bibinfo {title} {Beam dynamics in $\mathcal{P}\mathcal{T}$ symmetric optical lattices},\ }\href {https://doi.org/10.1103/PhysRevLett.100.103904} {\bibfield  {journal} {\bibinfo  {journal} {Phys. Rev. Lett.}\ }\textbf {\bibinfo {volume} {100}},\ \bibinfo {pages} {103904} (\bibinfo {year} {2008})}\BibitemShut {NoStop}%
\bibitem [{\citenamefont {Guo}\ \emph {et~al.}(2009)\citenamefont {Guo}, \citenamefont {Salamo}, \citenamefont {Duchesne}, \citenamefont {Morandotti}, \citenamefont {Volatier-Ravat}, \citenamefont {Aimez}, \citenamefont {Siviloglou},\ and\ \citenamefont {Christodoulides}}]{Guo2009}%
  \BibitemOpen
  \bibfield  {author} {\bibinfo {author} {\bibfnamefont {A.}~\bibnamefont {Guo}}, \bibinfo {author} {\bibfnamefont {G.~J.}\ \bibnamefont {Salamo}}, \bibinfo {author} {\bibfnamefont {D.}~\bibnamefont {Duchesne}}, \bibinfo {author} {\bibfnamefont {R.}~\bibnamefont {Morandotti}}, \bibinfo {author} {\bibfnamefont {M.}~\bibnamefont {Volatier-Ravat}}, \bibinfo {author} {\bibfnamefont {V.}~\bibnamefont {Aimez}}, \bibinfo {author} {\bibfnamefont {G.~A.}\ \bibnamefont {Siviloglou}},\ and\ \bibinfo {author} {\bibfnamefont {D.~N.}\ \bibnamefont {Christodoulides}},\ }\bibfield  {title} {\bibinfo {title} {Observation of $\mathcal{P}\mathcal{T}$-symmetry breaking in complex optical potentials},\ }\href {https://doi.org/10.1103/PhysRevLett.103.093902} {\bibfield  {journal} {\bibinfo  {journal} {Phys. Rev. Lett.}\ }\textbf {\bibinfo {volume} {103}},\ \bibinfo {pages} {093902} (\bibinfo {year} {2009})}\BibitemShut {NoStop}%
\bibitem [{\citenamefont {R{\"u}ter}\ \emph {et~al.}(2010)\citenamefont {R{\"u}ter}, \citenamefont {Makris}, \citenamefont {El-Ganainy}, \citenamefont {Christodoulides}, \citenamefont {Segev},\ and\ \citenamefont {Kip}}]{Ruter2010}%
  \BibitemOpen
  \bibfield  {author} {\bibinfo {author} {\bibfnamefont {C.~E.}\ \bibnamefont {R{\"u}ter}}, \bibinfo {author} {\bibfnamefont {K.~G.}\ \bibnamefont {Makris}}, \bibinfo {author} {\bibfnamefont {R.}~\bibnamefont {El-Ganainy}}, \bibinfo {author} {\bibfnamefont {D.~N.}\ \bibnamefont {Christodoulides}}, \bibinfo {author} {\bibfnamefont {M.}~\bibnamefont {Segev}},\ and\ \bibinfo {author} {\bibfnamefont {D.}~\bibnamefont {Kip}},\ }\bibfield  {title} {\bibinfo {title} {Observation of parity--time symmetry in optics},\ }\href {https://www.nature.com/articles/nphys1515} {\bibfield  {journal} {\bibinfo  {journal} {Nature Phys.}\ }\textbf {\bibinfo {volume} {6}},\ \bibinfo {pages} {192} (\bibinfo {year} {2010})}\BibitemShut {NoStop}%
\bibitem [{\citenamefont {El-Ganainy}\ \emph {et~al.}(2018)\citenamefont {El-Ganainy}, \citenamefont {Makris}, \citenamefont {Khajavikhan}, \citenamefont {Musslimani}, \citenamefont {Rotter},\ and\ \citenamefont {Christodoulides}}]{El-Ganainy2018}%
  \BibitemOpen
  \bibfield  {author} {\bibinfo {author} {\bibfnamefont {R.}~\bibnamefont {El-Ganainy}}, \bibinfo {author} {\bibfnamefont {K.~G.}\ \bibnamefont {Makris}}, \bibinfo {author} {\bibfnamefont {M.}~\bibnamefont {Khajavikhan}}, \bibinfo {author} {\bibfnamefont {Z.~H.}\ \bibnamefont {Musslimani}}, \bibinfo {author} {\bibfnamefont {S.}~\bibnamefont {Rotter}},\ and\ \bibinfo {author} {\bibfnamefont {D.~N.}\ \bibnamefont {Christodoulides}},\ }\bibfield  {title} {\bibinfo {title} {Non-{Hermitian} physics and {PT} symmetry},\ }\href {https://doi.org/10.1038/nphys4323} {\bibfield  {journal} {\bibinfo  {journal} {Nat. Phys.}\ }\textbf {\bibinfo {volume} {14}},\ \bibinfo {pages} {11} (\bibinfo {year} {2018})}\BibitemShut {NoStop}%
\bibitem [{\citenamefont {Poli}\ \emph {et~al.}(2015)\citenamefont {Poli}, \citenamefont {Bellec}, \citenamefont {Kuhl}, \citenamefont {Mortessagne},\ and\ \citenamefont {Schomerus}}]{Poli:2015}%
  \BibitemOpen
  \bibfield  {author} {\bibinfo {author} {\bibfnamefont {C.}~\bibnamefont {Poli}}, \bibinfo {author} {\bibfnamefont {M.}~\bibnamefont {Bellec}}, \bibinfo {author} {\bibfnamefont {U.}~\bibnamefont {Kuhl}}, \bibinfo {author} {\bibfnamefont {F.}~\bibnamefont {Mortessagne}},\ and\ \bibinfo {author} {\bibfnamefont {H.}~\bibnamefont {Schomerus}},\ }\bibfield  {title} {\bibinfo {title} {Selective enhancement of topologically induced interface states in a dielectric resonator chain},\ }\href {https://doi.org/10.1038/ncomms7710} {\bibfield  {journal} {\bibinfo  {journal} {Nat. Commun.}\ }\textbf {\bibinfo {volume} {6}},\ \bibinfo {pages} {6710} (\bibinfo {year} {2015})}\BibitemShut {NoStop}%
\bibitem [{\citenamefont {Midya}\ \emph {et~al.}(2018)\citenamefont {Midya}, \citenamefont {Zhao},\ and\ \citenamefont {Feng}}]{Midya2018}%
  \BibitemOpen
  \bibfield  {author} {\bibinfo {author} {\bibfnamefont {B.}~\bibnamefont {Midya}}, \bibinfo {author} {\bibfnamefont {H.}~\bibnamefont {Zhao}},\ and\ \bibinfo {author} {\bibfnamefont {L.}~\bibnamefont {Feng}},\ }\bibfield  {title} {\bibinfo {title} {{Non-Hermitian photonics promises exceptional topology of light}},\ }\href {https://www.nature.com/articles/s41467-018-05175-8} {\bibfield  {journal} {\bibinfo  {journal} {Nature Commun.}\ }\textbf {\bibinfo {volume} {9}},\ \bibinfo {pages} {2674} (\bibinfo {year} {2018})}\BibitemShut {NoStop}%
\bibitem [{\citenamefont {Shen}\ \emph {et~al.}(2018)\citenamefont {Shen}, \citenamefont {Zhen},\ and\ \citenamefont {Fu}}]{Shen:2018}%
  \BibitemOpen
  \bibfield  {author} {\bibinfo {author} {\bibfnamefont {H.}~\bibnamefont {Shen}}, \bibinfo {author} {\bibfnamefont {B.}~\bibnamefont {Zhen}},\ and\ \bibinfo {author} {\bibfnamefont {L.}~\bibnamefont {Fu}},\ }\bibfield  {title} {\bibinfo {title} {Topological band theory for non-{Hermitian} {Hamiltonians}},\ }\href {https://doi.org/10.1103/PhysRevLett.120.146402} {\bibfield  {journal} {\bibinfo  {journal} {Phys. Rev. Lett.}\ }\textbf {\bibinfo {volume} {120}},\ \bibinfo {pages} {146402} (\bibinfo {year} {2018})}\BibitemShut {NoStop}%
\bibitem [{\citenamefont {Kawabata}\ \emph {et~al.}(2019)\citenamefont {Kawabata}, \citenamefont {Shiozaki}, \citenamefont {Ueda},\ and\ \citenamefont {Sato}}]{Kawabata:2019}%
  \BibitemOpen
  \bibfield  {author} {\bibinfo {author} {\bibfnamefont {K.}~\bibnamefont {Kawabata}}, \bibinfo {author} {\bibfnamefont {K.}~\bibnamefont {Shiozaki}}, \bibinfo {author} {\bibfnamefont {M.}~\bibnamefont {Ueda}},\ and\ \bibinfo {author} {\bibfnamefont {M.}~\bibnamefont {Sato}},\ }\bibfield  {title} {\bibinfo {title} {Symmetry and topology in non-{Hermitian} physics},\ }\href {https://doi.org/10.1103/PhysRevX.9.041015} {\bibfield  {journal} {\bibinfo  {journal} {Phys. Rev. X}\ }\textbf {\bibinfo {volume} {9}},\ \bibinfo {pages} {041015} (\bibinfo {year} {2019})}\BibitemShut {NoStop}%
\bibitem [{\citenamefont {Ota}\ \emph {et~al.}(2020)\citenamefont {Ota}, \citenamefont {Takata}, \citenamefont {Ozawa}, \citenamefont {Amo}, \citenamefont {Jia}, \citenamefont {Kante}, \citenamefont {Notomi}, \citenamefont {Arakawa},\ and\ \citenamefont {Iwamoto}}]{Ota2020}%
  \BibitemOpen
  \bibfield  {author} {\bibinfo {author} {\bibfnamefont {Y.}~\bibnamefont {Ota}}, \bibinfo {author} {\bibfnamefont {K.}~\bibnamefont {Takata}}, \bibinfo {author} {\bibfnamefont {T.}~\bibnamefont {Ozawa}}, \bibinfo {author} {\bibfnamefont {A.}~\bibnamefont {Amo}}, \bibinfo {author} {\bibfnamefont {Z.}~\bibnamefont {Jia}}, \bibinfo {author} {\bibfnamefont {B.}~\bibnamefont {Kante}}, \bibinfo {author} {\bibfnamefont {M.}~\bibnamefont {Notomi}}, \bibinfo {author} {\bibfnamefont {Y.}~\bibnamefont {Arakawa}},\ and\ \bibinfo {author} {\bibfnamefont {S.}~\bibnamefont {Iwamoto}},\ }\bibfield  {title} {\bibinfo {title} {Active topological photonics},\ }\href {https://www.degruyter.com/document/doi/10.1515/nanoph-2019-0376/html} {\bibfield  {journal} {\bibinfo  {journal} {Nanophotonics}\ }\textbf {\bibinfo {volume} {9}},\ \bibinfo {pages} {547} (\bibinfo {year} {2020})}\BibitemShut {NoStop}%
\bibitem [{\citenamefont {Bergholtz}\ \emph {et~al.}(2021)\citenamefont {Bergholtz}, \citenamefont {Budich},\ and\ \citenamefont {Kunst}}]{Bergholtz:2021}%
  \BibitemOpen
  \bibfield  {author} {\bibinfo {author} {\bibfnamefont {E.~J.}\ \bibnamefont {Bergholtz}}, \bibinfo {author} {\bibfnamefont {J.~C.}\ \bibnamefont {Budich}},\ and\ \bibinfo {author} {\bibfnamefont {F.~K.}\ \bibnamefont {Kunst}},\ }\bibfield  {title} {\bibinfo {title} {{Exceptional topology of non-Hermitian systems}},\ }\href {https://doi.org/10.1103/RevModPhys.93.015005} {\bibfield  {journal} {\bibinfo  {journal} {Rev. Mod. Phys.}\ }\textbf {\bibinfo {volume} {93}},\ \bibinfo {pages} {015005} (\bibinfo {year} {2021})}\BibitemShut {NoStop}%
\bibitem [{\citenamefont {Ding}\ \emph {et~al.}(2022)\citenamefont {Ding}, \citenamefont {Fang},\ and\ \citenamefont {Ma}}]{Ding:2022}%
  \BibitemOpen
  \bibfield  {author} {\bibinfo {author} {\bibfnamefont {K.}~\bibnamefont {Ding}}, \bibinfo {author} {\bibfnamefont {C.}~\bibnamefont {Fang}},\ and\ \bibinfo {author} {\bibfnamefont {G.}~\bibnamefont {Ma}},\ }\bibfield  {title} {\bibinfo {title} {{Non-Hermitian topology and exceptional-point geometries}},\ }\href {https://www.nature.com/articles/s42254-022-00516-5} {\bibfield  {journal} {\bibinfo  {journal} {Nature Reviews Physics}\ }\textbf {\bibinfo {volume} {4}},\ \bibinfo {pages} {745} (\bibinfo {year} {2022})}\BibitemShut {NoStop}%
\bibitem [{\citenamefont {Okuma}\ and\ \citenamefont {Sato}(2023)}]{Okuma:2023}%
  \BibitemOpen
  \bibfield  {author} {\bibinfo {author} {\bibfnamefont {N.}~\bibnamefont {Okuma}}\ and\ \bibinfo {author} {\bibfnamefont {M.}~\bibnamefont {Sato}},\ }\bibfield  {title} {\bibinfo {title} {{Non-Hermitian topological phenomena: A review}},\ }\href {https://www.annualreviews.org/content/journals/10.1146/annurev-conmatphys-040521-033133} {\bibfield  {journal} {\bibinfo  {journal} {Annu. Rev. Condens. Matter Phys.}\ }\textbf {\bibinfo {volume} {14}},\ \bibinfo {pages} {83} (\bibinfo {year} {2023})}\BibitemShut {NoStop}%
\bibitem [{\citenamefont {Garrison}\ and\ \citenamefont {Wright}(1988)}]{Garrison1988}%
  \BibitemOpen
  \bibfield  {author} {\bibinfo {author} {\bibfnamefont {J.}~\bibnamefont {Garrison}}\ and\ \bibinfo {author} {\bibfnamefont {E.~M.}\ \bibnamefont {Wright}},\ }\bibfield  {title} {\bibinfo {title} {Complex geometrical phases for dissipative systems},\ }\href {https://www.sciencedirect.com/science/article/pii/037596018890905X?via%3Dihub} {\bibfield  {journal} {\bibinfo  {journal} {Phys. Lett. A}\ }\textbf {\bibinfo {volume} {128}},\ \bibinfo {pages} {177} (\bibinfo {year} {1988})}\BibitemShut {NoStop}%
\bibitem [{\citenamefont {Dattoli}\ \emph {et~al.}(1990)\citenamefont {Dattoli}, \citenamefont {Mignani},\ and\ \citenamefont {Torre}}]{Dattoli1990}%
  \BibitemOpen
  \bibfield  {author} {\bibinfo {author} {\bibfnamefont {G.}~\bibnamefont {Dattoli}}, \bibinfo {author} {\bibfnamefont {R.}~\bibnamefont {Mignani}},\ and\ \bibinfo {author} {\bibfnamefont {A.}~\bibnamefont {Torre}},\ }\bibfield  {title} {\bibinfo {title} {{Geometrical phase in the cyclic evolution of non-Hermitian systems}},\ }\href {https://iopscience.iop.org/article/10.1088/0305-4470/23/24/020} {\bibfield  {journal} {\bibinfo  {journal} {J. Phys. A}\ }\textbf {\bibinfo {volume} {23}},\ \bibinfo {pages} {5795} (\bibinfo {year} {1990})}\BibitemShut {NoStop}%
\bibitem [{\citenamefont {Mondrag{\'o}n}\ and\ \citenamefont {Hern{\'a}ndez}(1996)}]{Mondragon1996}%
  \BibitemOpen
  \bibfield  {author} {\bibinfo {author} {\bibfnamefont {A.}~\bibnamefont {Mondrag{\'o}n}}\ and\ \bibinfo {author} {\bibfnamefont {E.}~\bibnamefont {Hern{\'a}ndez}},\ }\bibfield  {title} {\bibinfo {title} {Berry phase of a resonant state},\ }\href {https://iopscience.iop.org/article/10.1088/0305-4470/29/10/032} {\bibfield  {journal} {\bibinfo  {journal} {J. Phys. A}\ }\textbf {\bibinfo {volume} {29}},\ \bibinfo {pages} {2567} (\bibinfo {year} {1996})}\BibitemShut {NoStop}%
\bibitem [{\citenamefont {Bliokh}(1999)}]{Bliokh1999}%
  \BibitemOpen
  \bibfield  {author} {\bibinfo {author} {\bibfnamefont {K.~Y.}\ \bibnamefont {Bliokh}},\ }\bibfield  {title} {\bibinfo {title} {The appearance of a geometric-type instability in dynamic systems with adiabatically varying parameters},\ }\href {https://iopscience.iop.org/article/10.1088/0305-4470/32/13/007} {\bibfield  {journal} {\bibinfo  {journal} {Journal of Physics A: Mathematical and General}\ }\textbf {\bibinfo {volume} {32}},\ \bibinfo {pages} {2551} (\bibinfo {year} {1999})}\BibitemShut {NoStop}%
\bibitem [{\citenamefont {Keck}\ \emph {et~al.}(2003)\citenamefont {Keck}, \citenamefont {Korsch},\ and\ \citenamefont {Mossmann}}]{Keck2003}%
  \BibitemOpen
  \bibfield  {author} {\bibinfo {author} {\bibfnamefont {F.}~\bibnamefont {Keck}}, \bibinfo {author} {\bibfnamefont {H.}~\bibnamefont {Korsch}},\ and\ \bibinfo {author} {\bibfnamefont {S.}~\bibnamefont {Mossmann}},\ }\bibfield  {title} {\bibinfo {title} {Unfolding a diabolic point: a generalized crossing scenario},\ }\href {https://iopscience.iop.org/article/10.1088/0305-4470/36/8/310} {\bibfield  {journal} {\bibinfo  {journal} {J. Phys. A}\ }\textbf {\bibinfo {volume} {36}},\ \bibinfo {pages} {2125} (\bibinfo {year} {2003})}\BibitemShut {NoStop}%
\bibitem [{\citenamefont {Nesterov}\ and\ \citenamefont {de~la Cruz}(2008)}]{Nesterov2008}%
  \BibitemOpen
  \bibfield  {author} {\bibinfo {author} {\bibfnamefont {A.~I.}\ \bibnamefont {Nesterov}}\ and\ \bibinfo {author} {\bibfnamefont {F.~A.}\ \bibnamefont {de~la Cruz}},\ }\bibfield  {title} {\bibinfo {title} {Complex magnetic monopoles, geometric phases and quantum evolution in the vicinity of diabolic and exceptional points},\ }\href {https://iopscience.iop.org/article/10.1088/1751-8113/41/48/485304} {\bibfield  {journal} {\bibinfo  {journal} {J. Phys. A}\ }\textbf {\bibinfo {volume} {41}},\ \bibinfo {pages} {485304} (\bibinfo {year} {2008})}\BibitemShut {NoStop}%
\bibitem [{\citenamefont {Liang}\ and\ \citenamefont {Huang}(2013)}]{Liang2013}%
  \BibitemOpen
  \bibfield  {author} {\bibinfo {author} {\bibfnamefont {S.-D.}\ \bibnamefont {Liang}}\ and\ \bibinfo {author} {\bibfnamefont {G.-Y.}\ \bibnamefont {Huang}},\ }\bibfield  {title} {\bibinfo {title} {{Topological invariance and global Berry phase in non-Hermitian systems}},\ }\href {https://doi.org/10.1103/PhysRevA.87.012118} {\bibfield  {journal} {\bibinfo  {journal} {Phys. Rev. A}\ }\textbf {\bibinfo {volume} {87}},\ \bibinfo {pages} {012118} (\bibinfo {year} {2013})}\BibitemShut {NoStop}%
\bibitem [{\citenamefont {Singhal}\ \emph {et~al.}(2023)\citenamefont {Singhal}, \citenamefont {Martello}, \citenamefont {Agrawal}, \citenamefont {Ozawa}, \citenamefont {Price},\ and\ \citenamefont {Gadway}}]{Singhal2023}%
  \BibitemOpen
  \bibfield  {author} {\bibinfo {author} {\bibfnamefont {Y.}~\bibnamefont {Singhal}}, \bibinfo {author} {\bibfnamefont {E.}~\bibnamefont {Martello}}, \bibinfo {author} {\bibfnamefont {S.}~\bibnamefont {Agrawal}}, \bibinfo {author} {\bibfnamefont {T.}~\bibnamefont {Ozawa}}, \bibinfo {author} {\bibfnamefont {H.}~\bibnamefont {Price}},\ and\ \bibinfo {author} {\bibfnamefont {B.}~\bibnamefont {Gadway}},\ }\bibfield  {title} {\bibinfo {title} {{Measuring the adiabatic non-Hermitian Berry phase in feedback-coupled oscillators}},\ }\href {https://doi.org/10.1103/PhysRevResearch.5.L032026} {\bibfield  {journal} {\bibinfo  {journal} {Phys. Rev. Res.}\ }\textbf {\bibinfo {volume} {5}},\ \bibinfo {pages} {L032026} (\bibinfo {year} {2023})}\BibitemShut {NoStop}%
\bibitem [{\citenamefont {Ozawa}\ and\ \citenamefont {Hayata}(2024)}]{OzawaHayata}%
  \BibitemOpen
  \bibfield  {author} {\bibinfo {author} {\bibfnamefont {T.}~\bibnamefont {Ozawa}}\ and\ \bibinfo {author} {\bibfnamefont {T.}~\bibnamefont {Hayata}},\ }\bibfield  {title} {\bibinfo {title} {Two-dimensional lattice with an imaginary magnetic field},\ }\href {https://doi.org/10.1103/PhysRevB.109.085113} {\bibfield  {journal} {\bibinfo  {journal} {Phys. Rev. B}\ }\textbf {\bibinfo {volume} {109}},\ \bibinfo {pages} {085113} (\bibinfo {year} {2024})}\BibitemShut {NoStop}%
\bibitem [{\citenamefont {Chalker}\ and\ \citenamefont {Mehlig}(1998)}]{Chalker:1998}%
  \BibitemOpen
  \bibfield  {author} {\bibinfo {author} {\bibfnamefont {J.~T.}\ \bibnamefont {Chalker}}\ and\ \bibinfo {author} {\bibfnamefont {B.}~\bibnamefont {Mehlig}},\ }\bibfield  {title} {\bibinfo {title} {Eigenvector statistics in non-{H}ermitian random matrix ensembles},\ }\href {https://doi.org/10.1103/PhysRevLett.81.3367} {\bibfield  {journal} {\bibinfo  {journal} {Phys. Rev. Lett.}\ }\textbf {\bibinfo {volume} {81}},\ \bibinfo {pages} {3367} (\bibinfo {year} {1998})}\BibitemShut {NoStop}%
\bibitem [{\citenamefont {{Petermann}}(1979)}]{Petermann:1979}%
  \BibitemOpen
  \bibfield  {author} {\bibinfo {author} {\bibfnamefont {K.}~\bibnamefont {{Petermann}}},\ }\bibfield  {title} {\bibinfo {title} {Calculated spontaneous emission factor for double-heterostructure injection lasers with gain-induced waveguiding},\ }\href {https://doi.org/10.1109/JQE.1979.1070064} {\bibfield  {journal} {\bibinfo  {journal} {IEEE J. Quantum Electron.}\ }\textbf {\bibinfo {volume} {15}},\ \bibinfo {pages} {566} (\bibinfo {year} {1979})}\BibitemShut {NoStop}%
\bibitem [{\citenamefont {Siegman}(1989)}]{Siegman:1989}%
  \BibitemOpen
  \bibfield  {author} {\bibinfo {author} {\bibfnamefont {A.~E.}\ \bibnamefont {Siegman}},\ }\bibfield  {title} {\bibinfo {title} {Excess spontaneous emission in non-{Hermitian} optical systems. {I.} {Laser} amplifiers},\ }\href {https://doi.org/10.1103/PhysRevA.39.1253} {\bibfield  {journal} {\bibinfo  {journal} {Phys. Rev. A}\ }\textbf {\bibinfo {volume} {39}},\ \bibinfo {pages} {1253} (\bibinfo {year} {1989})}\BibitemShut {NoStop}%
\bibitem [{\citenamefont {Patra}\ \emph {et~al.}(2000)\citenamefont {Patra}, \citenamefont {Schomerus},\ and\ \citenamefont {Beenakker}}]{Patra:2000}%
  \BibitemOpen
  \bibfield  {author} {\bibinfo {author} {\bibfnamefont {M.}~\bibnamefont {Patra}}, \bibinfo {author} {\bibfnamefont {H.}~\bibnamefont {Schomerus}},\ and\ \bibinfo {author} {\bibfnamefont {C.~W.~J.}\ \bibnamefont {Beenakker}},\ }\bibfield  {title} {\bibinfo {title} {Quantum-limited linewidth of a chaotic laser cavity},\ }\href {https://doi.org/10.1103/PhysRevA.61.023810} {\bibfield  {journal} {\bibinfo  {journal} {Phys. Rev. A}\ }\textbf {\bibinfo {volume} {61}},\ \bibinfo {pages} {023810} (\bibinfo {year} {2000})}\BibitemShut {NoStop}%
\bibitem [{\citenamefont {Heiss}(2000)}]{Heiss:2000}%
  \BibitemOpen
  \bibfield  {author} {\bibinfo {author} {\bibfnamefont {W.~D.}\ \bibnamefont {Heiss}},\ }\bibfield  {title} {\bibinfo {title} {Repulsion of resonance states and exceptional points},\ }\href {https://doi.org/10.1103/PhysRevE.61.929} {\bibfield  {journal} {\bibinfo  {journal} {Phys. Rev. E}\ }\textbf {\bibinfo {volume} {61}},\ \bibinfo {pages} {929} (\bibinfo {year} {2000})}\BibitemShut {NoStop}%
\bibitem [{\citenamefont {Schomerus}(2020)}]{Schomerus2020}%
  \BibitemOpen
  \bibfield  {author} {\bibinfo {author} {\bibfnamefont {H.}~\bibnamefont {Schomerus}},\ }\bibfield  {title} {\bibinfo {title} {Nonreciprocal response theory of non-{H}ermitian mechanical metamaterials: Response phase transition from the skin effect of zero modes},\ }\href {https://doi.org/10.1103/PhysRevResearch.2.013058} {\bibfield  {journal} {\bibinfo  {journal} {Phys. Rev. Res.}\ }\textbf {\bibinfo {volume} {2}},\ \bibinfo {pages} {013058} (\bibinfo {year} {2020})}\BibitemShut {NoStop}%
\bibitem [{\citenamefont {Berry}(2004)}]{Berry:2004}%
  \BibitemOpen
  \bibfield  {author} {\bibinfo {author} {\bibfnamefont {M.~V.}\ \bibnamefont {Berry}},\ }\bibfield  {title} {\bibinfo {title} {Physics of nonhermitian degeneracies},\ }\href {https://doi.org/10.1023/B:CJOP.0000044002.05657.04} {\bibfield  {journal} {\bibinfo  {journal} {Czech. J. Physics}\ }\textbf {\bibinfo {volume} {54}},\ \bibinfo {pages} {1039} (\bibinfo {year} {2004})}\BibitemShut {NoStop}%
\bibitem [{\citenamefont {Miri}\ and\ \citenamefont {Alu}(2019)}]{Miri2019}%
  \BibitemOpen
  \bibfield  {author} {\bibinfo {author} {\bibfnamefont {M.-A.}\ \bibnamefont {Miri}}\ and\ \bibinfo {author} {\bibfnamefont {A.}~\bibnamefont {Alu}},\ }\bibfield  {title} {\bibinfo {title} {Exceptional points in optics and photonics},\ }\href {https://www.science.org/doi/full/10.1126/science.aar7709} {\bibfield  {journal} {\bibinfo  {journal} {Science}\ }\textbf {\bibinfo {volume} {363}},\ \bibinfo {pages} {eaar7709} (\bibinfo {year} {2019})}\BibitemShut {NoStop}%
\bibitem [{\citenamefont {Yoo}\ \emph {et~al.}(2011)\citenamefont {Yoo}, \citenamefont {Sim},\ and\ \citenamefont {Schomerus}}]{Yoo:2011}%
  \BibitemOpen
  \bibfield  {author} {\bibinfo {author} {\bibfnamefont {G.}~\bibnamefont {Yoo}}, \bibinfo {author} {\bibfnamefont {H.-S.}\ \bibnamefont {Sim}},\ and\ \bibinfo {author} {\bibfnamefont {H.}~\bibnamefont {Schomerus}},\ }\bibfield  {title} {\bibinfo {title} {Quantum noise and mode nonorthogonality in non-{Hermitian} {PT}-symmetric optical resonators},\ }\href {https://doi.org/10.1103/PhysRevA.84.063833} {\bibfield  {journal} {\bibinfo  {journal} {Phys. Rev. A}\ }\textbf {\bibinfo {volume} {84}},\ \bibinfo {pages} {063833} (\bibinfo {year} {2011})}\BibitemShut {NoStop}%
\bibitem [{\citenamefont {Takata}\ \emph {et~al.}(2021)\citenamefont {Takata}, \citenamefont {Nozaki}, \citenamefont {Kuramochi}, \citenamefont {Matsuo}, \citenamefont {Takeda}, \citenamefont {Fujii}, \citenamefont {Kita}, \citenamefont {Shinya},\ and\ \citenamefont {Notomi}}]{Takata:2021}%
  \BibitemOpen
  \bibfield  {author} {\bibinfo {author} {\bibfnamefont {K.}~\bibnamefont {Takata}}, \bibinfo {author} {\bibfnamefont {K.}~\bibnamefont {Nozaki}}, \bibinfo {author} {\bibfnamefont {E.}~\bibnamefont {Kuramochi}}, \bibinfo {author} {\bibfnamefont {S.}~\bibnamefont {Matsuo}}, \bibinfo {author} {\bibfnamefont {K.}~\bibnamefont {Takeda}}, \bibinfo {author} {\bibfnamefont {T.}~\bibnamefont {Fujii}}, \bibinfo {author} {\bibfnamefont {S.}~\bibnamefont {Kita}}, \bibinfo {author} {\bibfnamefont {A.}~\bibnamefont {Shinya}},\ and\ \bibinfo {author} {\bibfnamefont {M.}~\bibnamefont {Notomi}},\ }\bibfield  {title} {\bibinfo {title} {Observing exceptional point degeneracy of radiation with electrically pumped photonic crystal coupled-nanocavity lasers},\ }\href {https://doi.org/10.1364/OPTICA.412596} {\bibfield  {journal} {\bibinfo  {journal} {Optica}\ }\textbf {\bibinfo {volume} {8}},\ \bibinfo {pages} {184} (\bibinfo {year} {2021})}\BibitemShut {NoStop}%
\bibitem [{\citenamefont {Hashemi}\ \emph {et~al.}(2022)\citenamefont {Hashemi}, \citenamefont {Busch}, \citenamefont {Christodoulides}, \citenamefont {Ozdemir},\ and\ \citenamefont {El-Ganainy}}]{Hashemi:2022}%
  \BibitemOpen
  \bibfield  {author} {\bibinfo {author} {\bibfnamefont {A.}~\bibnamefont {Hashemi}}, \bibinfo {author} {\bibfnamefont {K.}~\bibnamefont {Busch}}, \bibinfo {author} {\bibfnamefont {D.~N.}\ \bibnamefont {Christodoulides}}, \bibinfo {author} {\bibfnamefont {S.~K.}\ \bibnamefont {Ozdemir}},\ and\ \bibinfo {author} {\bibfnamefont {R.}~\bibnamefont {El-Ganainy}},\ }\bibfield  {title} {\bibinfo {title} {Linear response theory of open systems with exceptional points},\ }\href {https://doi.org/10.1038/s41467-022-30715-8} {\bibfield  {journal} {\bibinfo  {journal} {Nat. Commun.}\ }\textbf {\bibinfo {volume} {13}},\ \bibinfo {pages} {3281} (\bibinfo {year} {2022})}\BibitemShut {NoStop}%
\bibitem [{\citenamefont {Wiersig}(2014)}]{Wiersig:2014}%
  \BibitemOpen
  \bibfield  {author} {\bibinfo {author} {\bibfnamefont {J.}~\bibnamefont {Wiersig}},\ }\bibfield  {title} {\bibinfo {title} {Enhancing the sensitivity of frequency and energy splitting detection by using exceptional points: Application to microcavity sensors for single-particle detection},\ }\href {https://doi.org/10.1103/PhysRevLett.112.203901} {\bibfield  {journal} {\bibinfo  {journal} {Phys. Rev. Lett.}\ }\textbf {\bibinfo {volume} {112}},\ \bibinfo {pages} {203901} (\bibinfo {year} {2014})}\BibitemShut {NoStop}%
\bibitem [{\citenamefont {Chen}\ \emph {et~al.}(2017)\citenamefont {Chen}, \citenamefont {{Kaya {\"O}zdemir}}, \citenamefont {Zhao}, \citenamefont {Wiersig},\ and\ \citenamefont {Yang}}]{Chen:2017}%
  \BibitemOpen
  \bibfield  {author} {\bibinfo {author} {\bibfnamefont {W.}~\bibnamefont {Chen}}, \bibinfo {author} {\bibfnamefont {{\c{S}}.}~\bibnamefont {{Kaya {\"O}zdemir}}}, \bibinfo {author} {\bibfnamefont {G.}~\bibnamefont {Zhao}}, \bibinfo {author} {\bibfnamefont {J.}~\bibnamefont {Wiersig}},\ and\ \bibinfo {author} {\bibfnamefont {L.}~\bibnamefont {Yang}},\ }\bibfield  {title} {\bibinfo {title} {Exceptional points enhance sensing in an optical microcavity},\ }\href {https://doi.org/10.1038/nature23281} {\bibfield  {journal} {\bibinfo  {journal} {Nature}\ }\textbf {\bibinfo {volume} {548}},\ \bibinfo {pages} {192} (\bibinfo {year} {2017})}\BibitemShut {NoStop}%
\bibitem [{\citenamefont {Wiersig}(2020)}]{Wiersig2020}%
  \BibitemOpen
  \bibfield  {author} {\bibinfo {author} {\bibfnamefont {J.}~\bibnamefont {Wiersig}},\ }\bibfield  {title} {\bibinfo {title} {Review of exceptional point-based sensors},\ }\href {https://doi.org/10.1364/PRJ.396115} {\bibfield  {journal} {\bibinfo  {journal} {Photon. Res.}\ }\textbf {\bibinfo {volume} {8}},\ \bibinfo {pages} {1457} (\bibinfo {year} {2020})}\BibitemShut {NoStop}%
\bibitem [{\citenamefont {van Eijkelenborg}\ \emph {et~al.}(1996)\citenamefont {van Eijkelenborg}, \citenamefont {Lindberg}, \citenamefont {Thijssen},\ and\ \citenamefont {Woerdman}}]{vanEijkelenborg:1996}%
  \BibitemOpen
  \bibfield  {author} {\bibinfo {author} {\bibfnamefont {M.~A.}\ \bibnamefont {van Eijkelenborg}}, \bibinfo {author} {\bibfnamefont {A.~M.}\ \bibnamefont {Lindberg}}, \bibinfo {author} {\bibfnamefont {M.~S.}\ \bibnamefont {Thijssen}},\ and\ \bibinfo {author} {\bibfnamefont {J.~P.}\ \bibnamefont {Woerdman}},\ }\bibfield  {title} {\bibinfo {title} {Resonance of quantum noise in an unstable cavity laser},\ }\href {https://doi.org/10.1103/PhysRevLett.77.4314} {\bibfield  {journal} {\bibinfo  {journal} {Phys. Rev. Lett.}\ }\textbf {\bibinfo {volume} {77}},\ \bibinfo {pages} {4314} (\bibinfo {year} {1996})}\BibitemShut {NoStop}%
\bibitem [{\citenamefont {Cheng}\ \emph {et~al.}(1996)\citenamefont {Cheng}, \citenamefont {Fanning},\ and\ \citenamefont {Siegman}}]{Cheng:1996}%
  \BibitemOpen
  \bibfield  {author} {\bibinfo {author} {\bibfnamefont {Y.-J.}\ \bibnamefont {Cheng}}, \bibinfo {author} {\bibfnamefont {C.~G.}\ \bibnamefont {Fanning}},\ and\ \bibinfo {author} {\bibfnamefont {A.~E.}\ \bibnamefont {Siegman}},\ }\bibfield  {title} {\bibinfo {title} {Experimental observation of a large excess quantum noise factor in the linewidth of a laser oscillator having nonorthogonal modes},\ }\href {https://doi.org/10.1103/PhysRevLett.77.627} {\bibfield  {journal} {\bibinfo  {journal} {Phys. Rev. Lett.}\ }\textbf {\bibinfo {volume} {77}},\ \bibinfo {pages} {627} (\bibinfo {year} {1996})}\BibitemShut {NoStop}%
\bibitem [{\citenamefont {Nenciu}\ and\ \citenamefont {Rasche}(1992)}]{Nenciu:1992}%
  \BibitemOpen
  \bibfield  {author} {\bibinfo {author} {\bibfnamefont {G.}~\bibnamefont {Nenciu}}\ and\ \bibinfo {author} {\bibfnamefont {G.}~\bibnamefont {Rasche}},\ }\bibfield  {title} {\bibinfo {title} {{On the adiabatic theorem for nonself-adjoint Hamiltonians}},\ }\href {https://iopscience.iop.org/article/10.1088/0305-4470/25/21/027} {\bibfield  {journal} {\bibinfo  {journal} {J. Phys. A}\ }\textbf {\bibinfo {volume} {25}},\ \bibinfo {pages} {5741} (\bibinfo {year} {1992})}\BibitemShut {NoStop}%
\bibitem [{\citenamefont {H\"oller}\ \emph {et~al.}(2020)\citenamefont {H\"oller}, \citenamefont {Read},\ and\ \citenamefont {Harris}}]{Holler:2020}%
  \BibitemOpen
  \bibfield  {author} {\bibinfo {author} {\bibfnamefont {J.}~\bibnamefont {H\"oller}}, \bibinfo {author} {\bibfnamefont {N.}~\bibnamefont {Read}},\ and\ \bibinfo {author} {\bibfnamefont {J.~G.~E.}\ \bibnamefont {Harris}},\ }\bibfield  {title} {\bibinfo {title} {Non-{Hermitian} adiabatic transport in spaces of exceptional points},\ }\href {https://doi.org/10.1103/PhysRevA.102.032216} {\bibfield  {journal} {\bibinfo  {journal} {Phys. Rev. A}\ }\textbf {\bibinfo {volume} {102}},\ \bibinfo {pages} {032216} (\bibinfo {year} {2020})}\BibitemShut {NoStop}%
\bibitem [{\citenamefont {Graefe}\ \emph {et~al.}(2013)\citenamefont {Graefe}, \citenamefont {Mailybaev},\ and\ \citenamefont {Moiseyev}}]{Graefe:2013}%
  \BibitemOpen
  \bibfield  {author} {\bibinfo {author} {\bibfnamefont {E.-M.}\ \bibnamefont {Graefe}}, \bibinfo {author} {\bibfnamefont {A.~A.}\ \bibnamefont {Mailybaev}},\ and\ \bibinfo {author} {\bibfnamefont {N.}~\bibnamefont {Moiseyev}},\ }\bibfield  {title} {\bibinfo {title} {Breakdown of adiabatic transfer of light in waveguides in the presence of absorption},\ }\href {https://doi.org/10.1103/PhysRevA.88.033842} {\bibfield  {journal} {\bibinfo  {journal} {Phys. Rev. A}\ }\textbf {\bibinfo {volume} {88}},\ \bibinfo {pages} {033842} (\bibinfo {year} {2013})}\BibitemShut {NoStop}%
\bibitem [{\citenamefont {Silberstein}\ \emph {et~al.}(2020)\citenamefont {Silberstein}, \citenamefont {Behrends}, \citenamefont {Goldstein},\ and\ \citenamefont {Ilan}}]{Silberstein2020}%
  \BibitemOpen
  \bibfield  {author} {\bibinfo {author} {\bibfnamefont {N.}~\bibnamefont {Silberstein}}, \bibinfo {author} {\bibfnamefont {J.}~\bibnamefont {Behrends}}, \bibinfo {author} {\bibfnamefont {M.}~\bibnamefont {Goldstein}},\ and\ \bibinfo {author} {\bibfnamefont {R.}~\bibnamefont {Ilan}},\ }\bibfield  {title} {\bibinfo {title} {{Berry connection induced anomalous wave-packet dynamics in non-Hermitian systems}},\ }\href {https://doi.org/10.1103/PhysRevB.102.245147} {\bibfield  {journal} {\bibinfo  {journal} {Phys. Rev. B}\ }\textbf {\bibinfo {volume} {102}},\ \bibinfo {pages} {245147} (\bibinfo {year} {2020})}\BibitemShut {NoStop}%
\bibitem [{Note1()}]{Note1}%
  \BibitemOpen
  \bibinfo {note} {If $\protect \bm {\lambda }$ is not three-dimensional, we need to define the Berry curvature as a two-form via the Berry connection one-form by \begin {align} \DOTSB \sum@ \slimits@ _{i<j}\Omega ^{LR}_{ij} (\protect \bm {\lambda }) d\lambda _i \wedge d\lambda _j \equiv \DOTSB \sum@ \slimits@ _{i,j} \partial _{\lambda _i}\protect \mathcal {A}^{LR}_j (\protect \bm {\lambda }) d\lambda _i \wedge d\lambda _j. \end {align}}\BibitemShut {NoStop}%
\bibitem [{Note2()}]{Note2}%
  \BibitemOpen
  \bibinfo {note} {We note that, since $\langle L | R\rangle \protect \neq 0$ for left and right eigenstates of the same nondegenerate eigenvalue, $M|R\rangle $ never vanishes.}\BibitemShut {Stop}%
\bibitem [{\citenamefont {Ghatak}\ \emph {et~al.}(2020)\citenamefont {Ghatak}, \citenamefont {Brandenbourger}, \citenamefont {van Wezel},\ and\ \citenamefont {Coulais}}]{ghatak}%
  \BibitemOpen
  \bibfield  {author} {\bibinfo {author} {\bibfnamefont {A.}~\bibnamefont {Ghatak}}, \bibinfo {author} {\bibfnamefont {M.}~\bibnamefont {Brandenbourger}}, \bibinfo {author} {\bibfnamefont {J.}~\bibnamefont {van Wezel}},\ and\ \bibinfo {author} {\bibfnamefont {C.}~\bibnamefont {Coulais}},\ }\bibfield  {title} {\bibinfo {title} {{Observation of non-{H}ermitian topology and its bulk–edge correspondence in an active mechanical metamaterial}},\ }\href {https://doi.org/10.1073/pnas.2010580117} {\bibfield  {journal} {\bibinfo  {journal} {PNAS}\ }\textbf {\bibinfo {volume} {117}},\ \bibinfo {pages} {29561} (\bibinfo {year} {2020})}\BibitemShut {NoStop}%
\bibitem [{\citenamefont {Nassar}\ \emph {et~al.}(2018)\citenamefont {Nassar}, \citenamefont {Chen}, \citenamefont {Norris},\ and\ \citenamefont {Huang}}]{Nassar:2018}%
  \BibitemOpen
  \bibfield  {author} {\bibinfo {author} {\bibfnamefont {H.}~\bibnamefont {Nassar}}, \bibinfo {author} {\bibfnamefont {H.}~\bibnamefont {Chen}}, \bibinfo {author} {\bibfnamefont {A.~N.}\ \bibnamefont {Norris}},\ and\ \bibinfo {author} {\bibfnamefont {G.~L.}\ \bibnamefont {Huang}},\ }\bibfield  {title} {\bibinfo {title} {Quantization of band tilting in modulated phononic crystals},\ }\href {https://doi.org/10.1103/PhysRevB.97.014305} {\bibfield  {journal} {\bibinfo  {journal} {Phys. Rev. B}\ }\textbf {\bibinfo {volume} {97}},\ \bibinfo {pages} {014305} (\bibinfo {year} {2018})}\BibitemShut {NoStop}%
\bibitem [{\citenamefont {Stegmaier}\ \emph {et~al.}(2024)\citenamefont {Stegmaier}, \citenamefont {Brand}, \citenamefont {Imhof}, \citenamefont {Fritzsche}, \citenamefont {Helbig}, \citenamefont {Hofmann}, \citenamefont {Boettcher}, \citenamefont {Greiter}, \citenamefont {Lee}, \citenamefont {Bahl}, \citenamefont {Szameit}, \citenamefont {Kie\ss{}ling}, \citenamefont {Thomale},\ and\ \citenamefont {Upreti}}]{Stegmaier:2024}%
  \BibitemOpen
  \bibfield  {author} {\bibinfo {author} {\bibfnamefont {A.}~\bibnamefont {Stegmaier}}, \bibinfo {author} {\bibfnamefont {H.}~\bibnamefont {Brand}}, \bibinfo {author} {\bibfnamefont {S.}~\bibnamefont {Imhof}}, \bibinfo {author} {\bibfnamefont {A.}~\bibnamefont {Fritzsche}}, \bibinfo {author} {\bibfnamefont {T.}~\bibnamefont {Helbig}}, \bibinfo {author} {\bibfnamefont {T.}~\bibnamefont {Hofmann}}, \bibinfo {author} {\bibfnamefont {I.}~\bibnamefont {Boettcher}}, \bibinfo {author} {\bibfnamefont {M.}~\bibnamefont {Greiter}}, \bibinfo {author} {\bibfnamefont {C.~H.}\ \bibnamefont {Lee}}, \bibinfo {author} {\bibfnamefont {G.}~\bibnamefont {Bahl}}, \bibinfo {author} {\bibfnamefont {A.}~\bibnamefont {Szameit}}, \bibinfo {author} {\bibfnamefont {T.}~\bibnamefont {Kie\ss{}ling}}, \bibinfo {author} {\bibfnamefont {R.}~\bibnamefont {Thomale}},\ and\ \bibinfo {author} {\bibfnamefont {L.~K.}\ \bibnamefont {Upreti}},\ }\bibfield  {title} {\bibinfo {title} {Realizing efficient topological temporal pumping in electrical
  circuits},\ }\href {https://doi.org/10.1103/PhysRevResearch.6.023010} {\bibfield  {journal} {\bibinfo  {journal} {Phys. Rev. Res.}\ }\textbf {\bibinfo {volume} {6}},\ \bibinfo {pages} {023010} (\bibinfo {year} {2024})}\BibitemShut {NoStop}%
\bibitem [{\citenamefont {Anandwade}\ \emph {et~al.}(2023)\citenamefont {Anandwade}, \citenamefont {Singhal}, \citenamefont {Paladugu}, \citenamefont {Martello}, \citenamefont {Castle}, \citenamefont {Agrawal}, \citenamefont {Carlson}, \citenamefont {Battle-McDonald}, \citenamefont {Ozawa}, \citenamefont {Price},\ and\ \citenamefont {Gadway}}]{Anandwade2023}%
  \BibitemOpen
  \bibfield  {author} {\bibinfo {author} {\bibfnamefont {R.}~\bibnamefont {Anandwade}}, \bibinfo {author} {\bibfnamefont {Y.}~\bibnamefont {Singhal}}, \bibinfo {author} {\bibfnamefont {S.~N.~M.}\ \bibnamefont {Paladugu}}, \bibinfo {author} {\bibfnamefont {E.}~\bibnamefont {Martello}}, \bibinfo {author} {\bibfnamefont {M.}~\bibnamefont {Castle}}, \bibinfo {author} {\bibfnamefont {S.}~\bibnamefont {Agrawal}}, \bibinfo {author} {\bibfnamefont {E.}~\bibnamefont {Carlson}}, \bibinfo {author} {\bibfnamefont {C.}~\bibnamefont {Battle-McDonald}}, \bibinfo {author} {\bibfnamefont {T.}~\bibnamefont {Ozawa}}, \bibinfo {author} {\bibfnamefont {H.~M.}\ \bibnamefont {Price}},\ and\ \bibinfo {author} {\bibfnamefont {B.}~\bibnamefont {Gadway}},\ }\bibfield  {title} {\bibinfo {title} {Synthetic mechanical lattices with synthetic interactions},\ }\href {https://doi.org/10.1103/PhysRevA.108.012221} {\bibfield  {journal} {\bibinfo  {journal} {Phys. Rev. A}\ }\textbf {\bibinfo {volume} {108}},\ \bibinfo {pages} {012221} (\bibinfo
  {year} {2023})}\BibitemShut {NoStop}%
\bibitem [{\citenamefont {Budich}\ and\ \citenamefont {Bergholtz}(2020)}]{Budich2020}%
  \BibitemOpen
  \bibfield  {author} {\bibinfo {author} {\bibfnamefont {J.~C.}\ \bibnamefont {Budich}}\ and\ \bibinfo {author} {\bibfnamefont {E.~J.}\ \bibnamefont {Bergholtz}},\ }\bibfield  {title} {\bibinfo {title} {Non-{H}ermitian topological sensors},\ }\href {https://doi.org/10.1103/PhysRevLett.125.180403} {\bibfield  {journal} {\bibinfo  {journal} {Phys. Rev. Lett.}\ }\textbf {\bibinfo {volume} {125}},\ \bibinfo {pages} {180403} (\bibinfo {year} {2020})}\BibitemShut {NoStop}%
\bibitem [{\citenamefont {Wanjura}\ \emph {et~al.}(2020)\citenamefont {Wanjura}, \citenamefont {Brunelli},\ and\ \citenamefont {Nunnenkamp}}]{Wanjura2020}%
  \BibitemOpen
  \bibfield  {author} {\bibinfo {author} {\bibfnamefont {C.~C.}\ \bibnamefont {Wanjura}}, \bibinfo {author} {\bibfnamefont {M.}~\bibnamefont {Brunelli}},\ and\ \bibinfo {author} {\bibfnamefont {A.}~\bibnamefont {Nunnenkamp}},\ }\bibfield  {title} {\bibinfo {title} {Topological framework for directional amplification in driven-dissipative cavity arrays},\ }\href {https://doi.org/10.1038/s41467-020-16863-9} {\bibfield  {journal} {\bibinfo  {journal} {Nat. Commun.}\ }\textbf {\bibinfo {volume} {11}},\ \bibinfo {pages} {3149} (\bibinfo {year} {2020})}\BibitemShut {NoStop}%
\bibitem [{\citenamefont {McDonald}\ and\ \citenamefont {Clerk}(2020)}]{McDonald2020}%
  \BibitemOpen
  \bibfield  {author} {\bibinfo {author} {\bibfnamefont {A.}~\bibnamefont {McDonald}}\ and\ \bibinfo {author} {\bibfnamefont {A.~A.}\ \bibnamefont {Clerk}},\ }\bibfield  {title} {\bibinfo {title} {Exponentially-enhanced quantum sensing with non-{Hermitian} lattice dynamics},\ }\href {https://doi.org/10.1038/s41467-020-19090-4} {\bibfield  {journal} {\bibinfo  {journal} {Nat. Commun.}\ }\textbf {\bibinfo {volume} {11}},\ \bibinfo {pages} {5382} (\bibinfo {year} {2020})}\BibitemShut {NoStop}%
\bibitem [{\citenamefont {Okuma}\ \emph {et~al.}(2020)\citenamefont {Okuma}, \citenamefont {Kawabata}, \citenamefont {Shiozaki},\ and\ \citenamefont {Sato}}]{Okuma2020}%
  \BibitemOpen
  \bibfield  {author} {\bibinfo {author} {\bibfnamefont {N.}~\bibnamefont {Okuma}}, \bibinfo {author} {\bibfnamefont {K.}~\bibnamefont {Kawabata}}, \bibinfo {author} {\bibfnamefont {K.}~\bibnamefont {Shiozaki}},\ and\ \bibinfo {author} {\bibfnamefont {M.}~\bibnamefont {Sato}},\ }\bibfield  {title} {\bibinfo {title} {Topological origin of non-{H}ermitian skin effects},\ }\href {https://doi.org/10.1103/PhysRevLett.124.086801} {\bibfield  {journal} {\bibinfo  {journal} {Phys. Rev. Lett.}\ }\textbf {\bibinfo {volume} {124}},\ \bibinfo {pages} {086801} (\bibinfo {year} {2020})}\BibitemShut {NoStop}%
\bibitem [{\citenamefont {Hofmann}\ \emph {et~al.}(2020)\citenamefont {Hofmann}, \citenamefont {Helbig}, \citenamefont {Schindler}, \citenamefont {Salgo}, \citenamefont {Brzezi\ifmmode~\acute{n}\else \'{n}\fi{}ska}, \citenamefont {Greiter}, \citenamefont {Kiessling}, \citenamefont {Wolf}, \citenamefont {Vollhardt}, \citenamefont {Kaba\ifmmode~\check{s}\else \v{s}\fi{}i}, \citenamefont {Lee}, \citenamefont {Bilu\ifmmode \check{s}\else \v{s}\fi{}i\ifmmode~\acute{c}\else \'{c}\fi{}}, \citenamefont {Thomale},\ and\ \citenamefont {Neupert}}]{Hofmann2020}%
  \BibitemOpen
  \bibfield  {author} {\bibinfo {author} {\bibfnamefont {T.}~\bibnamefont {Hofmann}}, \bibinfo {author} {\bibfnamefont {T.}~\bibnamefont {Helbig}}, \bibinfo {author} {\bibfnamefont {F.}~\bibnamefont {Schindler}}, \bibinfo {author} {\bibfnamefont {N.}~\bibnamefont {Salgo}}, \bibinfo {author} {\bibfnamefont {M.}~\bibnamefont {Brzezi\ifmmode~\acute{n}\else \'{n}\fi{}ska}}, \bibinfo {author} {\bibfnamefont {M.}~\bibnamefont {Greiter}}, \bibinfo {author} {\bibfnamefont {T.}~\bibnamefont {Kiessling}}, \bibinfo {author} {\bibfnamefont {D.}~\bibnamefont {Wolf}}, \bibinfo {author} {\bibfnamefont {A.}~\bibnamefont {Vollhardt}}, \bibinfo {author} {\bibfnamefont {A.}~\bibnamefont {Kaba\ifmmode~\check{s}\else \v{s}\fi{}i}}, \bibinfo {author} {\bibfnamefont {C.~H.}\ \bibnamefont {Lee}}, \bibinfo {author} {\bibfnamefont {A.}~\bibnamefont {Bilu\ifmmode \check{s}\else \v{s}\fi{}i\ifmmode~\acute{c}\else \'{c}\fi{}}}, \bibinfo {author} {\bibfnamefont {R.}~\bibnamefont {Thomale}},\ and\ \bibinfo {author} {\bibfnamefont
  {T.}~\bibnamefont {Neupert}},\ }\bibfield  {title} {\bibinfo {title} {Reciprocal skin effect and its realization in a topolectrical circuit},\ }\href {https://doi.org/10.1103/PhysRevResearch.2.023265} {\bibfield  {journal} {\bibinfo  {journal} {Phys. Rev. Res.}\ }\textbf {\bibinfo {volume} {2}},\ \bibinfo {pages} {023265} (\bibinfo {year} {2020})}\BibitemShut {NoStop}%
\bibitem [{\citenamefont {Kane}\ and\ \citenamefont {Lubensky}(2014)}]{Kane:2014}%
  \BibitemOpen
  \bibfield  {author} {\bibinfo {author} {\bibfnamefont {C.~L.}\ \bibnamefont {Kane}}\ and\ \bibinfo {author} {\bibfnamefont {T.~C.}\ \bibnamefont {Lubensky}},\ }\bibfield  {title} {\bibinfo {title} {Topological boundary modes in isostatic lattices},\ }\href {https://doi.org/10.1038/nphys2835} {\bibfield  {journal} {\bibinfo  {journal} {Nature Phys.}\ }\textbf {\bibinfo {volume} {10}},\ \bibinfo {pages} {39} (\bibinfo {year} {2014})}\BibitemShut {NoStop}%
\bibitem [{\citenamefont {Chen}\ \emph {et~al.}(2014)\citenamefont {Chen}, \citenamefont {Upadhyaya},\ and\ \citenamefont {Vitelli}}]{Chen:2014}%
  \BibitemOpen
  \bibfield  {author} {\bibinfo {author} {\bibfnamefont {B.~G.}\ \bibnamefont {Chen}}, \bibinfo {author} {\bibfnamefont {N.}~\bibnamefont {Upadhyaya}},\ and\ \bibinfo {author} {\bibfnamefont {V.}~\bibnamefont {Vitelli}},\ }\bibfield  {title} {\bibinfo {title} {Nonlinear conduction via solitons in a topological mechanical insulator},\ }\href {https://doi.org/10.1073/pnas.1405969111} {\bibfield  {journal} {\bibinfo  {journal} {PNAS}\ }\textbf {\bibinfo {volume} {111}},\ \bibinfo {pages} {13004} (\bibinfo {year} {2014})}\BibitemShut {NoStop}%
\bibitem [{\citenamefont {Huber}(2016)}]{Huber:2016}%
  \BibitemOpen
  \bibfield  {author} {\bibinfo {author} {\bibfnamefont {S.~D.}\ \bibnamefont {Huber}},\ }\bibfield  {title} {\bibinfo {title} {Topological mechanics},\ }\href {https://doi.org/10.1038/nphys3801} {\bibfield  {journal} {\bibinfo  {journal} {Nature Phys.}\ }\textbf {\bibinfo {volume} {12}},\ \bibinfo {pages} {621} (\bibinfo {year} {2016})}\BibitemShut {NoStop}%
\bibitem [{Note3()}]{Note3}%
  \BibitemOpen
  \bibinfo {note} {Since $\protect \mathcal {Q}$ is $N \times N^\prime $ and the rank of $\protect \mathcal {Q}$ is $N^\prime $, there exists an $N^\prime \times N$ left-inverse matrix of $\protect \mathcal {Q}$, which we denote by $\protect \mathcal {Q}_\protect \mathrm {left}^{-1}$, satisfying $\protect \mathcal {Q}_\protect \mathrm {left}^{-1} \protect \mathcal {Q} = \protect \mathbb {I}_{N^\prime }$. Here, $\protect \mathbb {I}_{N^\prime }$ is an $N^\prime \times N^\prime $ identity matrix. Using this left-inverse matrix, we define $\protect \mathbf {p}$ by $\protect \mathbf {p} \equiv \protect \mathcal {Q}_\protect \mathrm {left}^{-1} \protect \frac {d}{dt}\protect \mathbf {x}$. It is then straightforward to show that the first-order differential equation~(\ref {eq:eom2}) is satisfied.}\BibitemShut {Stop}%
\bibitem [{Note4()}]{Note4}%
  \BibitemOpen
  \bibinfo {note} {For $\protect \mathcal {Q}^T=\protect \mathcal {R}$ the effective Hamiltonian can be brought into symmetric form by a unitary transformation $UHU^{-1}$, where $U=\protect \mathrm {diag}(\protect \mathbb {I}_{N},i\protect \mathbb {I}_{N'})$. This condition matches the definition of reciprocity in ordinary mechanics, where $\protect \mathcal {Q}$ and $\protect \mathcal {R}$ are furthermore real. The mathematical alternative $\protect \mathcal {Q}^T=-\protect \mathcal {R}$ results, for real $Q$ and $R$, in a negative semi-definite matrix $M$, which corresponds to a mechanically unstable system.}\BibitemShut {Stop}%
\bibitem [{\citenamefont {Beenakker}(1998)}]{Beenakker:1998}%
  \BibitemOpen
  \bibfield  {author} {\bibinfo {author} {\bibfnamefont {C.~W.~J.}\ \bibnamefont {Beenakker}},\ }\bibfield  {title} {\bibinfo {title} {Thermal radiation and amplified spontaneous emission from a random medium},\ }\href {https://doi.org/10.1103/PhysRevLett.81.1829} {\bibfield  {journal} {\bibinfo  {journal} {Phys. Rev. Lett.}\ }\textbf {\bibinfo {volume} {81}},\ \bibinfo {pages} {1829} (\bibinfo {year} {1998})}\BibitemShut {NoStop}%
\bibitem [{\citenamefont {Aharonov}\ and\ \citenamefont {Anandan}(1987)}]{Aharonov1987}%
  \BibitemOpen
  \bibfield  {author} {\bibinfo {author} {\bibfnamefont {Y.}~\bibnamefont {Aharonov}}\ and\ \bibinfo {author} {\bibfnamefont {J.}~\bibnamefont {Anandan}},\ }\bibfield  {title} {\bibinfo {title} {Phase change during a cyclic quantum evolution},\ }\href {https://doi.org/10.1103/PhysRevLett.58.1593} {\bibfield  {journal} {\bibinfo  {journal} {Phys. Rev. Lett.}\ }\textbf {\bibinfo {volume} {58}},\ \bibinfo {pages} {1593} (\bibinfo {year} {1987})}\BibitemShut {NoStop}%
\end{thebibliography}%

\end{document}